\documentclass[12pt]{article}

\usepackage{color}
\usepackage{graphics}
\usepackage{graphicx}
\usepackage{scicite}
\usepackage{times}
\usepackage{amsmath,amssymb}
\usepackage{siunitx}
\usepackage{tabularx}
\usepackage[normalem]{ulem}
\usepackage[utf8x]{inputenc} % for non-ASCII characters
\usepackage[font=small]{caption}

\usepackage{float} %<--- AM; added this 17.10.2019

\newenvironment{sciabstract}{
\begin{quote} \bf}
{\end{quote}}

\title{Sub-picosecond thermalization dynamics in condensation of strongly coupled lattice plasmons}

\author
{Aaro I. V\"{a}kev\"{a}inen,$^{1}$ Antti J. Moilanen,$^{1}$ Marek Ne\v cada,$^{1}$ \\ Tommi K. Hakala,$^{2}$ Konstantinos S. Daskalakis,$^{1}$ P\"{a}ivi T\"{o}rm\"{a}$^{1\ast}$\\
\\
\normalsize{$^{1}$Department of Applied Physics, Aalto University School of Science,}\\
\normalsize{P.O. Box 15100, Aalto, FI-00076, Finland}\\
\normalsize{$^{2}$Institute  of  Photonics, University  of  Eastern  Finland,}\\
\normalsize{P.O. Box 111, Joensuu, FI-80101, Finland}\\
\\
\normalsize{$^\ast$To whom correspondence should be addressed; E-mail:  paivi.torma@aalto.fi. }
}

\date{}

\begin{document} 

% Double-space the manuscript.
\baselineskip20pt
\linespread{1.3}

% Make the title.
\maketitle

% Place your abstract within the special {sciabstract} environment.
%%%%%%%%%%%%%%%%%% ABSTRACT
\begin{sciabstract}

Bosonic condensates offer exciting prospects for studies of non-equilibrium quantum dynamics. Understanding the dynamics is particularly challenging in the sub-picosecond timescales typical for room temperature luminous driven-dissipative condensates. Here we combine a lattice of plasmonic nanoparticles with dye molecule solution at the strong coupling regime, and pump the molecules optically. The emitted light reveals three distinct regimes: one-dimensional lasing, incomplete stimulated thermalization, and two-dimensional multimode condensation. The condensate is achieved by matching the thermalization rate with the lattice size and occurs only for pump pulse durations below a critical value. Our results give access to control and monitoring of thermalization processes and condensate formation at sub-picosecond timescale.

\end{sciabstract}

\clearpage
\noindent
%\textbf{One sentence summary:} We show that a thermalized bosonic condensate can form in a 200~fs timescale due to stimulated processes and strong coupling. 
%\TKH{References to be used}
%\cite{Ramezani2018}
%\cite{KirtonPRA2015}
%\cite{radonjic_interplay_2018} %this was already in bib file
%\cite{weill_bose-einstein_2019}
%\cite{Damm2016}

%%%%%%%%%%%%%%%%%% INTRODUCTION
\section*{Introduction}

Ideal Bose-Einstein condensation means accumulation of macroscopic population to a single ground state in an equilibrium system, with emergence of long-range order. Bosonic condensation in non- or quasi-equilibrium and driven-dissipative systems extends this concept and offers plentiful new phenomena, such as loss of algebraically decaying phase order~\cite{altman_two-dimensional_2015,keeling_superfluidity_2017}, generalized Bose-Einstein condensation (BEC) into multiple states~\cite{vorberg_generalized_2013}, rich phase diagrams of lasing, condensation and superradiance phenomena~\cite{kirton_superradiant_2018,hesten_decondensation_2018}, and quantum simulation of the XY model that is at the heart of many optimization problems~\cite{berloff_realizing_2017}. Such condensates may also be a powerful system to explore dynamical quantum phase transitions~\cite{heyl_dynamical_2018}. Each presently available condensate system offers different advantages and limitations concerning studies of non- and quasi-equilibrium dynamics. The ability to tune interactions precisely over a wide range is the major advantage of ultracold gases~\cite{bloch_many-body_2008,torma_quantum_2015}. Polariton condensates~\cite{deng_condensation_2002,kasprzak_bose-einstein_2006,balili_bose-einstein_2007,baumberg_spontaneous_2008,kena-cohen_room-temperature_2010,carusotto_quantum_2013,byrnes_exciton-polariton_2014,daskalakis_nonlinear_2014,plumhof_room-temperature_2014} offer high critical temperatures compared to ultracold gases. In photon condensates~\cite{klaers_bose-einstein_2010,schmitt_dynamics_2018}, the thermal bath is easily controlled~\cite{schmitt_observation_2014,Schmitt2016PRL}. Recently periodic two-dimensional (2D) arrays of metal nanoparticles, so-called plasmonic lattices or crystals~\cite{wang_rich_2018}, have emerged as a multifaceted platform for room temperature lasing and condensation at  weak~\cite{zhou_lasing_2013,schokker_lasing_2014,yang_real-time_2015,hakala_lasing_2017,hakala_bose-einstein_2018} and strong coupling~\cite{ramezani_plasmon-exciton-polariton_2017,de_giorgi_interaction_2018,Ramezani2018} regimes.

In plasmonic lattices, the lattice geometry and periodicity, the size and shape of the nanoparticles, and the overall size of the lattice can be controlled with nanometer accuracy and independent of one another. The energy where condensation or lasing occurs is given by the band edge energy that depends on the period of the array. Remarkably the band edge energy and the dispersion are extremely constant over large lattices (accuracy 0.1\%~\cite{hakala_bose-einstein_2018}). In semiconductor polariton condensates, disorder in the samples often leads to traps and fragmentation~\cite{carusotto_quantum_2013}, or condensates may be trapped by geometry~\cite{klaers_bose-einstein_2010,byrnes_exciton-polariton_2014}. Thus plasmonic lattices offer a feature complementary to other condensate systems, namely that propagation of excitations over the lattice can be used for monitoring time-evolution of such processes as thermalization: each position in the array can be related to time via the group velocity, and there are no spurious effects due to non-uniformity of the sample. Spatially resolved luminescence was utilized in this way in the first observation of a BEC in a plasmonic lattice~\cite{hakala_bose-einstein_2018}.  

Here we show that formation of a condensate with a pronounced thermal distribution is possible at a 200~fs timescale and attribute this strikingly fast thermalization to partially coherent dynamics due to stimulated processes and strong coupling. We observe a unique double threshold phenomenon where one-dimensional (1D) lasing occurs for lower pump fluences and two-dimensional (2D) multimode condensation, associated with thermalization, at higher fluences. 
The transition between lasing and condensation shown in our work is different from previous condensates~\cite{byrnes_exciton-polariton_2014,bajoni_photon_2007,schmitt_thermalization_2015,Damm2017NatComm,walker_driven-dissipative_2018,hakala_bose-einstein_2018,weill_bose-einstein_2019}: it relies on matching the system size, propagation of excitations, and the thermalization dynamics. Importantly we find a peculiar intermediate regime showing features of a thermalization process but no macroscopic population at the lowest energy states. This regime allows us to reveal the stimulated nature of the thermalization process by the behavior of the luminescence in lattices of different sizes. As a direct evidence of the ultrafast character of the thermalization and condensation process, we show that it occurs only for pump pulse durations below a critical value of 100$-$250~fs. In the following, we first present characterization, such as luminescence spectra and spatial coherence, of the lasing and condensation phenomena and then focus on the main results: the stimulated nature of the thermalization process and the dramatic effect of the pump pulse duration.

\subsection*{System}

Our system consists of cylindrical gold nanoparticles in a rectangular lattice overlaid with a solution of organic dye molecule IR-792 (see details in Methods). The lattice supports dispersive modes, so-called surface lattice resonances (SLRs), which are hybrid modes composed of localized surface plasmon resonances at the nanoparticles and the diffracted orders of the periodic structure~\cite{wang_rich_2018,kravets_plasmonic_2018}. The electric field of the SLR modes is confined to the lattice plane in which the SLR excitations can propagate. An SLR excitation can be considered a bosonic quasiparticle that consists (mostly) of a photon and of collective electron oscillation in individual metal particles.

\begin{figure}[H]
  \centering
    \includegraphics[width=0.9\textwidth]{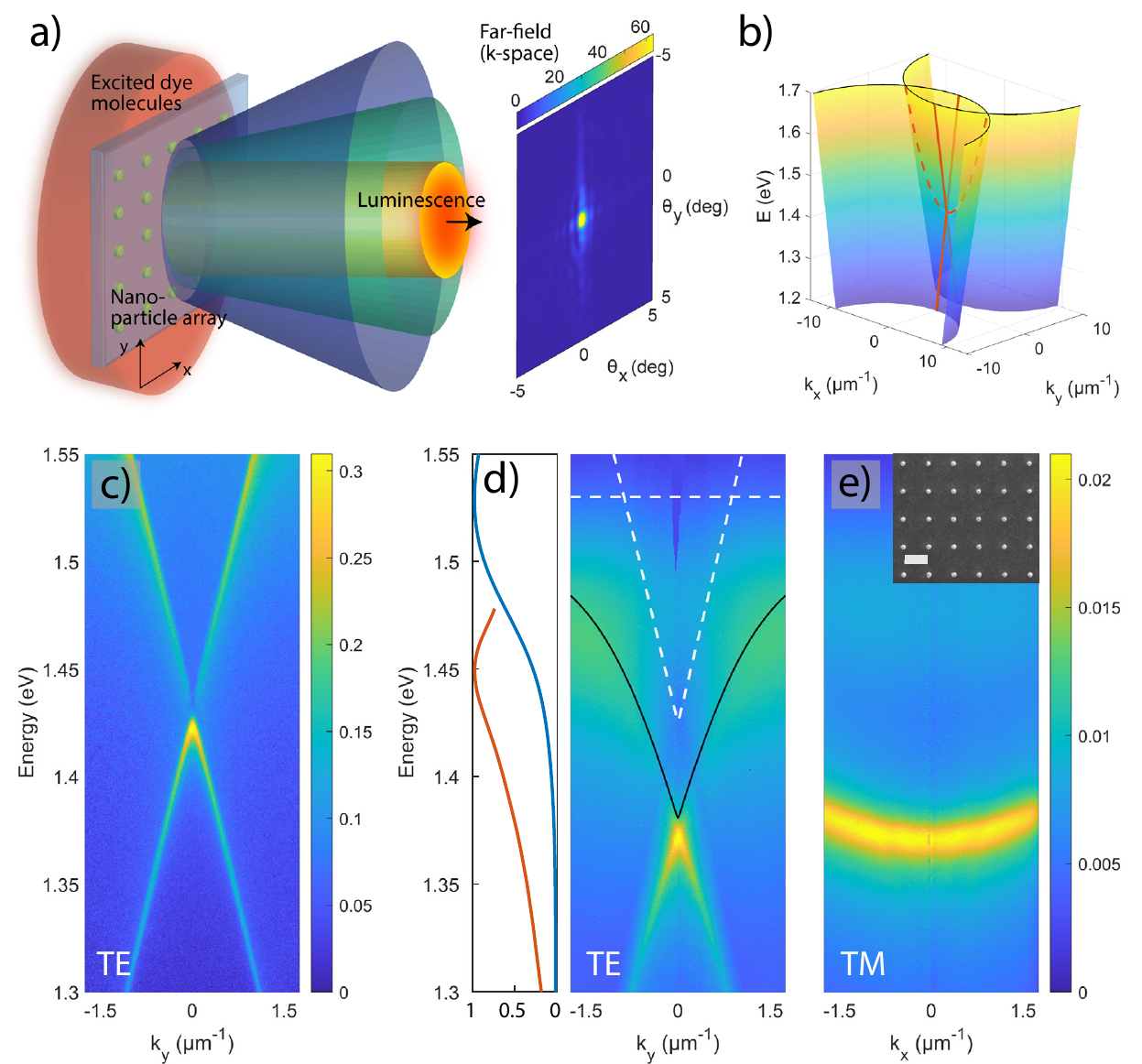}
   \caption{\textbf{Schematic of the system and dispersion of the modes.} (a) Artistic illustration of the experimental configuration. 
   (b) Light cones for the diffracted orders (0,-1) and (0,+1) that arise for $x$-polarized nanoparticles. Crosscut along $k_y$ ($k_x=0$) is called the TE mode (red solid lines) and along $k_x$ ($k_y=0$) the TM mode (red dashed line). For the TE mode, the polarization is perpendicular to the propagation direction ($\textbf{e}_x$,$k_y$), and for the TM mode parallel ($\textbf{e}_x$,$k_x$). Crosscuts of the SLR dispersions are experimentally obtained by measuring the transmission (c) without the molecule or (d-e) reflection with 80~mM solution of IR-792. In (d), the SLR dispersion exhibits a red shift and an avoided crossing with the absorption transition of the molecule. The absorption line and the uncoupled SLR mode are depicted with white dashed lines and the lower-polariton branch given by a coupled-modes-model fit is shown with a black line. Absorption and emission spectra of IR-792 are displayed in the left panel with blue and red lines, respectively. Scanning electron micrograph of the nanoparticle array is shown as an inset in (e), the scale bar is 500~nm.
  }
\end{figure}

The SLR modes are classified to transverse magnetic (TM) or transverse electric (TE) depending on the polarization and propagation direction, as defined in Figure~1b. The measured dispersions are displayed in Figure~1c-e. In the presence of dye molecules, the SLR dispersion shifts downwards in energy with respect to the initial diffracted order crossing. Moreover the TE modes begin to bend when approaching the molecular absorption line at 1.53~eV. These observations indicate strong coupling between the SLR and molecular excitations~\cite{torma_strong_2015}; a coupled modes fit to the data gives a Rabi splitting of 164~meV (larger than the average line width of the molecule (150~meV) and SLR (10~meV)) and an exciton part of 23\% at $k=0$. In the following, we refer to the hybrids of the SLR mode excitations and molecular excitons as polaritons, for brevity. The coherence length of the polaritons (samples that are not pumped) is 24 $\mu$m, as obtained from the observed dispersions.
To our best knowledge, this is the first time that high molecule concentration in a liquid gain solution is implemented in plasmonic lattice systems, cf. \cite{yang_real-time_2015,ramezani_plasmon-exciton-polariton_2017}. 
The plasmonic lattice, optimized for creating the condensate, has a particle diameter of 100~nm and height of 50~nm, the period in $y$- and $x$-direction of $p_y$ = 570~nm and $p_x$ = 620~nm, dye concentration of 80~mM, and a lattice size of 100$\times$\SI{100}{\micro m^2}. The period $p_y$ is varied between 520 and 590~nm, and the lattice size between 40$\times$40 and 200$\times$\SI{200}{\micro m^2}. 

We excite the sample with laser pulses at 1~kHz repetition rate and central wavelength of 800~nm, and resolve the luminescence spectrally as a function of angle and spatial position on the array, Figure~1a (for details see Methods). The pump does not directly couple to the SLR modes, and only a small fraction of the photons are coupled to the single particle resonance and/or absorbed by molecules within a near vicinity of the nanoparticle lattice. Active region of the dye molecules lies within a few hundred nanometers from the lattice plane, shown experimentally in~\cite{hakala_lasing_2017,daskalakis_ultrafast_2018,yang_unidirectional_2015,wang_band-edge_2017}, and molecules further away are unlikely to couple to the SLR modes. 

\section*{Results}

We study the luminescence properties of the plasmonic lattice as a function of pump fluence, that is, the energy per unit area per excitation pulse, and find a prominent double threshold behaviour. The system is excited with an $x$-polarized 50~fs laser pulse that has a flat intensity profile and a size larger than the lattice. Excited polariton modes continuously leak through radiative loss, and therefore the observed luminescence intensity is directly proportional to the population of the polaritons. We record real space and momentum ($k$-)space intensity distributions and the corresponding spectra, the photon energy is $E = hc/\lambda_0$ and the in-plane wave vector $k_{x,y} = 2\pi/\lambda_0 sin(\theta_{x,y})$. 

\begin{figure}[ht!]
\label{figure2}
  \centering
    \includegraphics[width=1.0\textwidth]{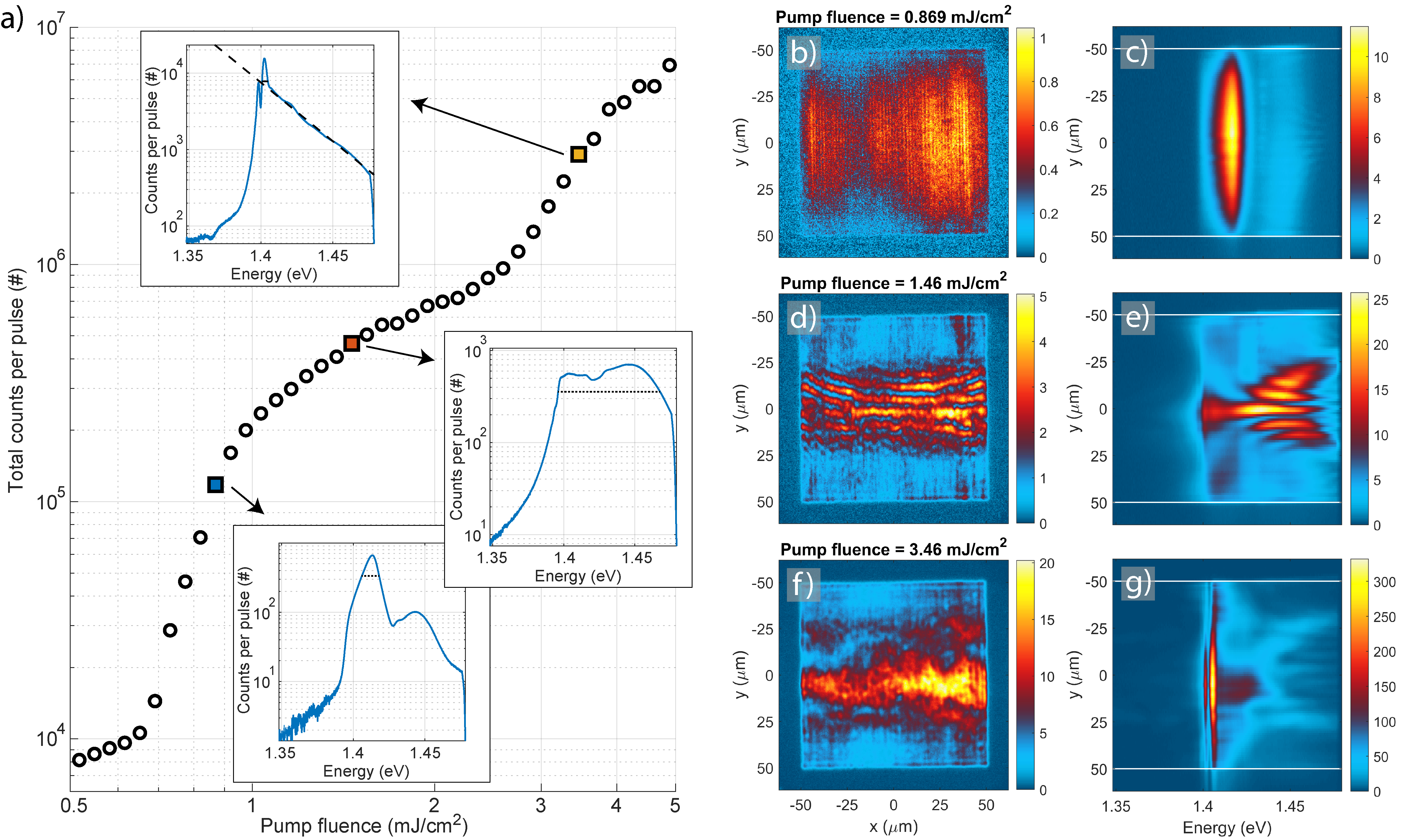}
   \caption{\textbf{Pump fluence dependence, real space images and spectra.} (a) Double-threshold curve of the pump fluence dependence of the total luminescence intensity. Insets: Line spectra obtained by integrating over the real space spectra in the $y$-direction (between the white lines). The FWHM of the spectral peaks is marked in the insets: 12~meV, 72~meV and 4.0~meV with increasing pump fluence. The dashed line in the top inset is a fit of the tail of the distribution to the Maxwell--Boltzmann distribution which gives a temperature of 313$\pm$2~K (95\% confidence bounds). 
   (b-g) Left column: Real space images of the plasmonic lattice. Right column: Spectral information of the luminescence as a function of $y$-position. The pump fluence is: (b-c) 0.87~mJcm$^{-2}$, (d-e) 1.5~mJcm$^{-2}$, (f-g) 3.5~mJcm$^{-2}$. The wave fronts in (d-e) arise from standing waves related to the momentum of the modes, for more information see Supplementary Figure~1 and Note~1. The real space images are recorded for single pump pulses but the corresponding spectra are integrated over 500 (c,e) and 70 pulses (g). 
  }
\end{figure}

%\AM{Comparison of real space images recorded for single pulses and integrated over 10 pulses is presented in Supplementary Figure 10.}

The sample luminescence as a function of pump fluence is presented in Figure~2. The total luminescence intensity reveals two non-linear thresholds and a linear intermediate regime, shown in Figure~2a. The line spectra, shown as insets, are obtained by integrating the real space spectra along the $y$-axis and unveil the population of polaritons as a function of energy. 
At the first threshold, Figure~2b-c, lasing (or polariton lasing/polariton condensation) typical for nanoparticle arrays~\cite{zhou_lasing_2013,hakala_lasing_2017,ramezani_plasmon-exciton-polariton_2017,de_giorgi_interaction_2018} is observed throughout the array. Increasing the pump fluence beyond the first threshold, Figure~2d-e, the luminescence becomes more intense in the central part compared to the top and bottom parts of the array. Moreover luminescence at the center takes place at a lower energy than at regions closer to the edges. We interpret this red shift as a signature of polariton population undergoing a thermalization process and propagating along the array in $+y$ and $-y$ directions, discussed below. At the second threshold, Figure~2f-g, the system undergoes a transition into a condensate (that presents a Maxwell-Boltzmann distribution at higher energies): the real space intensity distribution shows uniform luminescence in the central part of the array, and the line spectrum (Figure~2a, top inset) has a narrow peak at the band edge and a long thermalized tail at  higher energies. A fit of the tail to the Maxwell--Boltzmann distribution (dashed line) gives the room temperature, $T = $313$\pm$2~K (see Methods for more information).
We use the existence of a long (ranging over several $kT$, where $k$ is the Boltzmann constant) tail that fits the Maxwell--Boltzmann distribution as the criterion for distinguishing between what we call the condensate and lasing regimes. By the term condensate we thus refer here to the existence of a MB tail in addition to narrow peaked population at low energy; peaked population alone, without the tail, is referred to as lasing. Since we are at the strong coupling regime, the lasing that we see is actually polariton lasing which in the literature is called also polariton condensation~\cite{Deng2010,byrnes_exciton-polariton_2014}; we refer to this just by the words lasing regime, for brevity. As will be shown below, also the momentum-space confinement and spatial coherence properties of our condensate and lasing regimes differ dramatically, further confirming that the two phenomena are distinct.

The spectrometer counts per emitted condensate pulse correspond to a photon number of $\sim$10$^9$ (see Methods), which is roughly $10^5$ times more than in the first BEC in a plasmonic lattice~\cite{hakala_bose-einstein_2018}. The increased luminescence is attributed to stimulated processes and differences in the sample as well as the pump and detection geometry (Supplementary Note~2). This tremendous improvement has increased the signal-to-noise ratio so that we can now observe a prominent thermal tail (it is likely to be even longer but the data is cut due to filtering out the pump pulse) which is an important signature of the efficiency of the thermalization process even in ultrafast timescales. The upgrade of luminescence intensity is also crucial for future fundamental studies and applications of this type of condensate. For instance, thermodynamic quantities can be determined using the observed photon distribution~\cite{Damm2016}. To produce a condensate, the periodicity must be tuned with respect to the thermalization rate and the array size (Supplementary Note~1).

Three distinct regimes are also observed in the $k$-space intensity distributions. Figure~3 presents the $k$-space images and spectra for the same sample and pump parameters as in Figure~2. Figure~3a-c shows that lasing spreads in the TM mode to large $k$ (i.e., large emission angles), whereas thermalization of the polaritons occurs mostly along the TE mode, see Figure~3d-f. At the condensation threshold, Figure~3g-i, we observe confinement in both $k_x$ and $k_y$, implying that the condensate has a 2D nature, in contrast to the lasing regime where confinement is observed only in $k_y$. Figure~3j shows line spectra obtained by integrating along $k_y$ from TE mode crosscuts (Figure~3c,f,i); the spectrometer slit width of \SI{500}{\micro m} corresponds to $\pm 1.3^{\circ}$ around $\theta_x=0$ in the 2D $k$-space images. 
Intriguingly, at the condensation threshold, we see multiple modes highly occupied at $k_y$ = 0. The line spectrum at 3.49~mJcm$^{-2}$ shows three narrow peaks followed by a thermalized tail. The full-width at half-maximum (FWHM) of the highest peak is 3.3~meV, significantly narrower than the bare SLR mode (10~meV). The spacing of the multiple peaks decreases towards lower energy, ruling out the possibility of a trivial Fabry--Pérot interference. We have also observed that the spacing does not depend on the periodicity or lattice size in a straighforward manner, and is not caused by waveguide effects~\cite{schokker_lasing_2014,winkler_dual-wavelength_2020}.
Based on T-matrix simulations, we find that the cylindrical shape and finite size of the nanoparticles leads to three distinct modes around the $\Gamma$-point ($k_y=k_x=0$) energy in an infinite lattice, one of these modes being highly degenerate. This highlights the role of the nanoparticles beyond providing a periodic structure. The degeneracy is lifted by a finite lattice size, producing further distinct modes. The lattice size thus provides an additional means to tune the mode structure. For more information, see Supplementary Figure~2 and Note~3. While at thermal equilibrium condensation occurs to the lowest energy, in driven-dissipative systems condensation to several modes at distinct energies is possible~\cite{vorberg_generalized_2013,hesten_decondensation_2018}. Since we observe a temporally integrated signal, we cannot rule out the possibility of a single mode condensate temporally evolving between different states in the sub-picosecond scale.
 
\begin{figure}[H]
\label{figure3}
  \centering
    \includegraphics[width=1.0\textwidth]{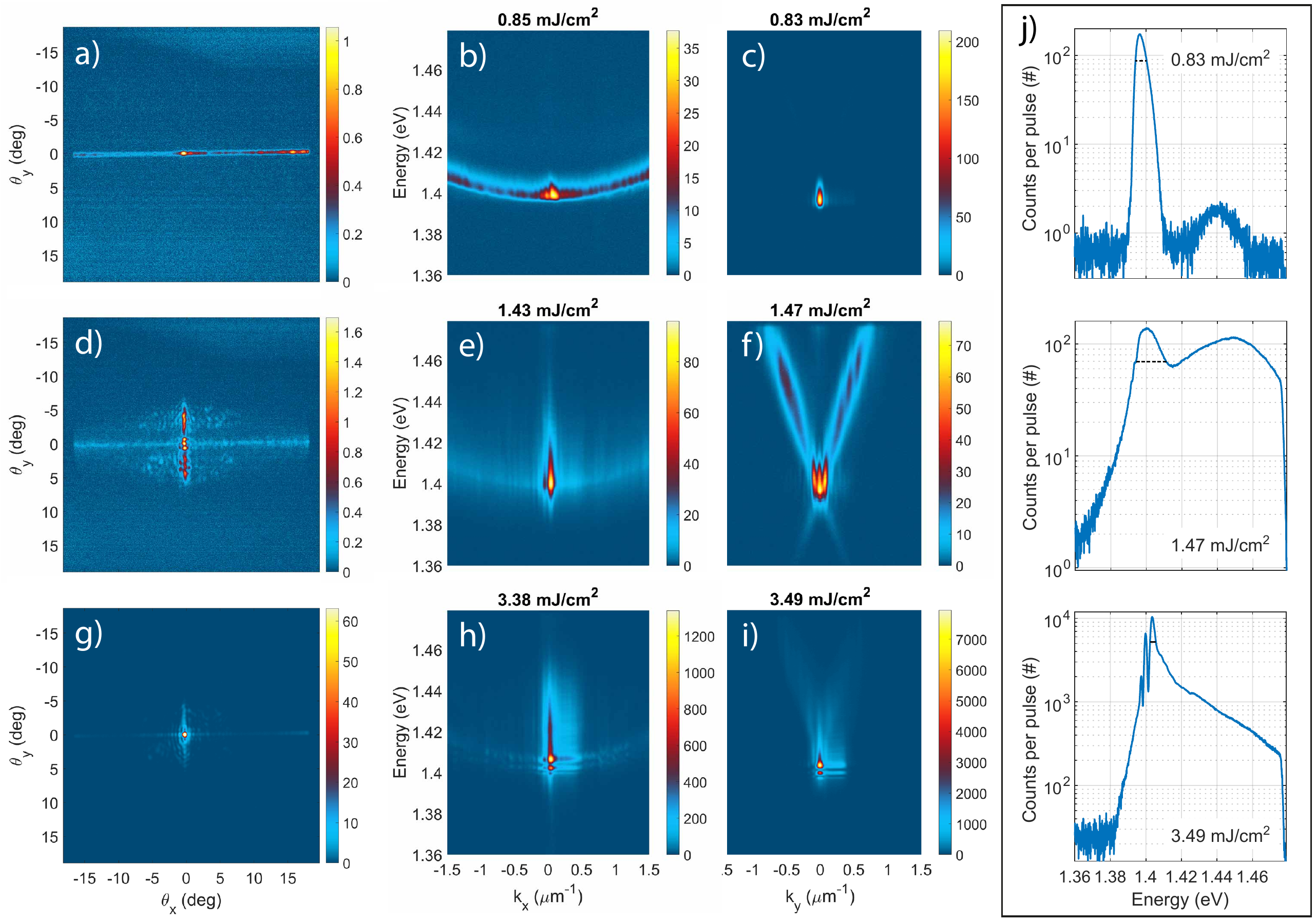}
   \caption{\textbf{Momentum space images and spectra.} First column: 2D momentum ($k$-)space images. Second and third column: $k$-space spectrum in the TM and TE mode directions, respectively. The spectra of the TM and TE modes correspond to horizontal and vertical slices of the 2D $k$-space spectrum, respectively. The pump fluence is: (a-c) 0.85~mJcm$^{-2}$, (d-f) 1.5~mJcm$^{-2}$, (g-i) 3.5~mJcm$^{-2}$, as in Figure~2. The images and the corresponding line spectra are integrated over 500 (c), 330 (f), and 20 pulses (i). The horizontal stretching of the emission peaks in (h-i) are CCD blooming artifacts. (j) Population distribution in the TE mode integrated from (c,f,i) along $k_y$. FWHMs are indicated with dashed black lines: 6.3~meV, 17~meV and 3.3~meV with increasing pump fluence. 
  }
\end{figure}
% 100 (b), 500 (c), 100 (e), 330 (f), and 20 pulses (h-i)

Lasing and condensation transitions are expected to result in increased spatial and temporal coherence of the emitted light. Increase in temporal coherence was evidenced as the narrowing of spectral line width (Figure~3j). To study the spatial coherence, we have performed a Michelson interferometer experiment as a function of pump fluence.  
In the Michelson interferometer, the real space image is split into two, one image is inverted and combined with the other one at the camera pixel array. The contrast of the observed interference fringes is extracted with a Fourier analysis of the spatial frequencies in the interfered image to exclude any artifacts produced by intensity variations in a single non-interfered image (Supplementary Figures~3-4 and Note~4).

In Figure~4, the fringe contrast (proportional to the first-order correlation function $g^{(1)}$) is shown as a function of pump fluence in both $y$- and $x$-directions of the lattice. High spatial coherence occurs in the $y$-direction over the array in both the lasing and condensation regimes (Figure~4b-c). At the intermediate regime, the spatial coherence decreases, in agreement with the observation that the thermalizing population dominates the luminescence signal. In the $x$-direction, spatial coherence is lower than in the $y$-direction over the whole pump range. However, the condensate clearly exhibits high spatial coherence also in $x$-direction, in contrast to lasing, which shows separated emission stripes in the real space image (see Figure~4d and Supplementary Figure~3). 
The spatial coherence measurements are in line with the observations from the 2D $k$-space images (Figure~3), where 1) lasing exhibits confined luminescence along $k_y$ (the direction of feedback) but spreads along $k_x$, 2) condensation shows 2D $k$-space confinement. The coherence both in the lasing and condensation cases extends over the whole array (100 $\mu$m x 100 $\mu$m), thus the coherence length is at least four times larger than that of the samples without pumping (24 $\mu$m). In a future study, larger samples should be used for finding how the coherence decays (e.g., exponential, polynomial). Whether algebraically decaying phase order exists in 2D driven-dissipative systems is a subtle question~\cite{altman_two-dimensional_2015,keeling_superfluidity_2017,zamora_tuning_2017,comaron_dynamical_2018}.

The lasing and condensation take place at energies 1.397 and 1.403~eV, respectively, lower than the band edge energy of the system without molecules, 1.423~eV. However the energies are blue-shifted from the lower polariton energy 1.373~eV (band edge in reflection) or 1.382~eV (fitted lower polariton branch; Figure~1d), which are experimentally obtained by a reflection measurement and coupled modes model~\cite{torma_strong_2015}. Since the whole dispersion gradually blue shifts as a function of pump fluence, it may be associated to degradation of strong coupling instead of originating from Coulomb interactions. Such saturation-caused non-linearity can also be considered as effective polariton-polariton interaction~\cite{carusotto_quantum_2013} (a rough estimate of the strength of such interaction in our case is given in Supplementary Note 8). Based on the band-edge locations (1.423$-$1.403)~eV~/~(1.423$-$1.373)~eV, the coupling in the condensation regime has decreased to $\sim$40\% of the case without pumping. Note that the observed double-threshold behaviour is different from semiconductor microcavity polariton condensates where condensation has a lower threshold than photonic lasing associated with loss of strong coupling~\cite{byrnes_exciton-polariton_2014,bajoni_photon_2007}.

\begin{figure}[H]
  \centering
    \includegraphics[width=0.95\textwidth]{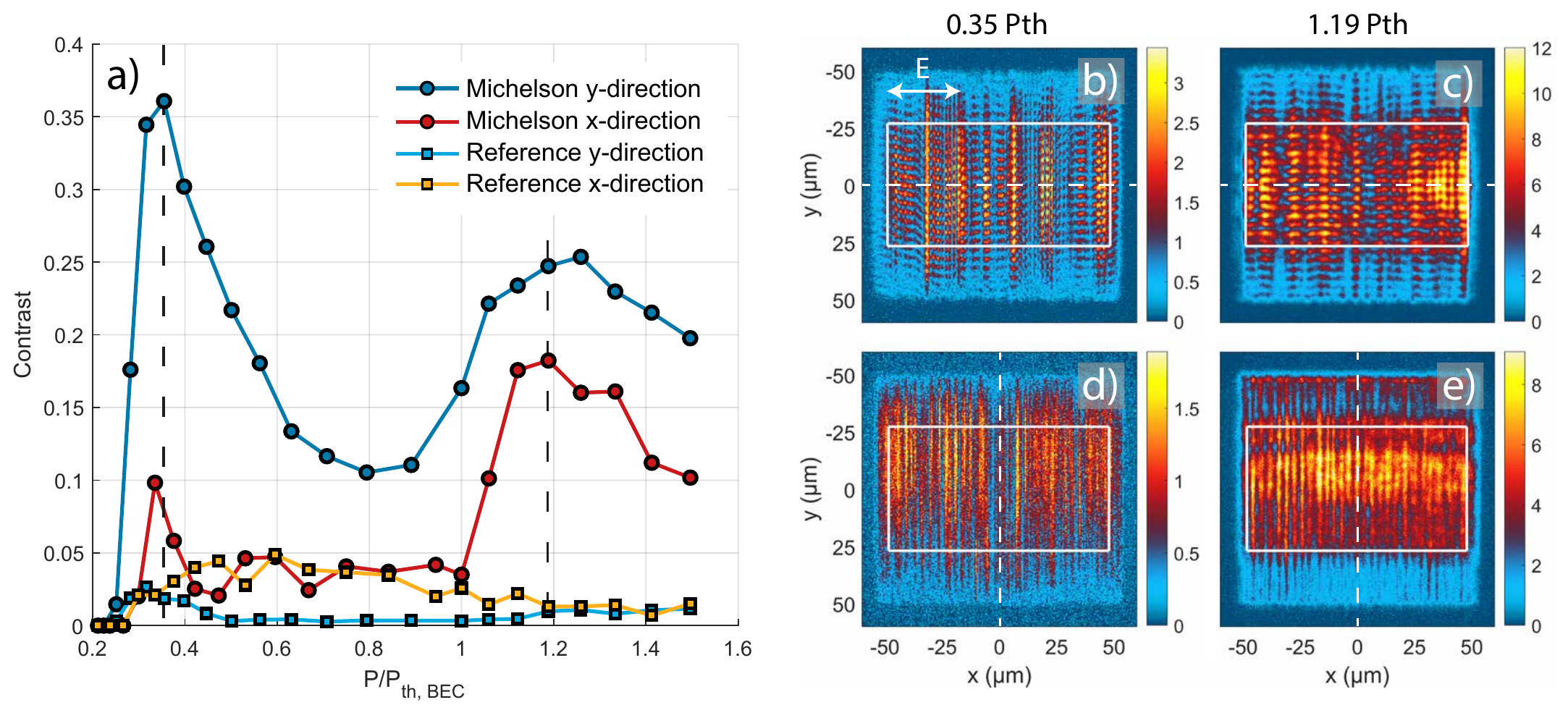}
   \caption{\textbf{Spatial coherence measurement along $y$ and $x$-axis of the plasmonic lattice.} (a) The Michelson interferometer fringe contrast averaged over the region of interest (white boxes in (b-e)) as a function of pump fluence. For reference, the contrast is extracted also from non-interfered real space images, as light blue and yellow curves. (b-e) Interfered real space images, where one of the images is inverted with respect to the white dashed line, to obtain the spatial coherence between $-y$ and $+y$ positions (b-c) or $-x$ and $+x$ positions (d-e). The interference patterns are shown for pump fluences corresponding to the lasing and condensation regimes (dashed lines in (a)). To obtain the spatial coherence between $-x$ and $+x$ positions, instead of rotating the detection optics, we rotate the sample (the lattice) and pump polarization by 90$^\circ$. The vertical stripes in b) and d) are due to one-dimensional lasing in y-direction, while the horizontal stripes (fringes) in b)-c) and vertical in e) arise from spatial coherence in the overlapped images in the Michelson interferometer. For more details see Methods and Supplementary Note~4. The real space images are recorded for single pump pulses.
   }
\end{figure}

\subsection*{Stimulated nature of the thermalization}

At the intermediate regime, we observed red shift of the luminescence as a function of distance in $y$-direction, Figure~2e. The trails of the red shift begin from the emission maximum of the dye molecule ($\sim$1.46~eV), at a certain distance from the array edges, and reach the band edge energy ($\sim$1.40~eV) exactly at the center of the 100$\times$\SI{100}{\micro m^2} array. 

To understand the red shift, we have recorded real space images and spatially-resolved spectra for different lattice sizes at intermediate pump fluences sufficient to trigger the thermalization process, Figure~5a-f. 
In a large array (Figure~5-e-f), we observe the trails of the red shift toward the center of the array, similarly as in the 100$\times$\SI{100}{\micro m^2} array (Figure~2d-e), but the red-shifting populations do not merge at the center.
In a small array (Figure~5a-b), the situation looks different at first glance since the red shift seems to occur from the center of the array toward the edges. However, careful comparison of the distance between the array edge and the location where the red shift begins (see Supplementary Figure~10 for details) reveals that for all arrays, for the given pump fluence, the distance is the same ($\sim$\SI{25}{\micro m} for 2.2~mJcm$^{-2}$ pump fluence presented in Figure~5a,c,e).

We explain this distance by stimulated emission pulse build-up time: the time between the maxima of the population inversion and the output pulse as defined in the rate equation simulation in Figure~5g (see~Ref.~\cite{daskalakis_ultrafast_2018} and Supplementary Note~5 for description of the model; note that we use this model just to illustrate the concept of pulse build-up in the thermalization process, not to describe the condensate or lasing observations). 
Pulse build-up is a well-known phenomenon in Q-switched lasers~\cite{siegman_lasers_1986}. In our system, the pump pulse excites the molecules, and the polaritons begin to propagate when the first photons populate the modes. The modes then gather gain while propagating, and therefore the peak of the stimulated emission pulse appears after a certain distance travelled along the array. This distance is seen as the dark zones in real space measurements, and it corresponds to the pulse build-up time. 
Note that this pulse is different to a lasing or condensate pulse since the line width of the luminescence is large and spatial coherence is small. 
By summing up such spatial intensity profiles of the thermalizing pulses at every point along the lattice (Figure~5h), we can reconstruct the real space intensity distributions (insets of Figure~5a,c,e). 
The dark zones appear because the edges do not receive propagating excitations from outside the lattice; the dark zones at the edges have approximately half the intensity of the central part. In the small lattice, intensity at the edges is similar to that of the larger lattices but in the center it is only half of that. 
Moreover the wavy interference patterns in the central part (Figure~5c,e) indeed only appear for arrays larger than 40$\times$\SI{40}{\micro m^2}, where there are counter-propagating pulses that can interfere.  

We found that the width of the dark zones depends on pump fluence as predicted by the rate-equation simulation and the theory of Q-switched lasers, namely the build-up time is inversely proportional to the pump fluence. Figure~5i shows that the dark-zone width follows the inverse of the pump fluence until it saturates at around 3~mJcm$^{-2}$ (corresponding to the BEC threshold) to a value below \SI{20}{\micro m} ($\sim$100$-$140~fs). The inset in Figure~5i shows the pulse build-up time extracted from our rate-equation simulation, and it displays a similar $\sim$1/P dependence.

We attribute the red shift to a thermalization process. At the intermediate regime of pump fluences, however, a thermal distribution is not reached before the population decays. For higher fluences, a condensate peak and a tail with Maxwell-Boltzmann (MB) distributed population emerges. A classical thermal MB distribution in logarithmic scale is a straight line. In contrast, a peaked feature at low energies, together with a MB tail, is a characteristic of the Bose-Einstein (BE) distribution. At low energies ($E - \mu < kT$), the BE distribution can be approximated as $kT/(E-\mu)$. A distribution of this form appears in so-called classical condensation of waves (or Rayleigh-Jeans (RJ) condensation), resulting for instance from an interplay between random noise and suitable gain/loss profiles of an optical system~\cite{SunNatPhys2012,Fischer2013,OrenOptica2014,RuckriegelPRL2014,KirtonPRA2015}. Our system does not have the conditions typical for classical condensation and, most importantly, the observed linear-in-log-scale tail does not match a distribution of the form $kT/(E-\mu)$. To rule out the RJ condensation by the distribution, one needs to observe it for energies $E$ that are larger than the condensate peak energy by more than $kT$, because for $E-\mu < kT$ the BE and RJ distributions coincide. It is thus essential that we resolve the tail up to energies that are $75$ meV above the condensate energy – three times larger than the room temperature $kT=25$ meV.

\begin{figure}[H]
  \centering
    \includegraphics[width=1.0\textwidth]{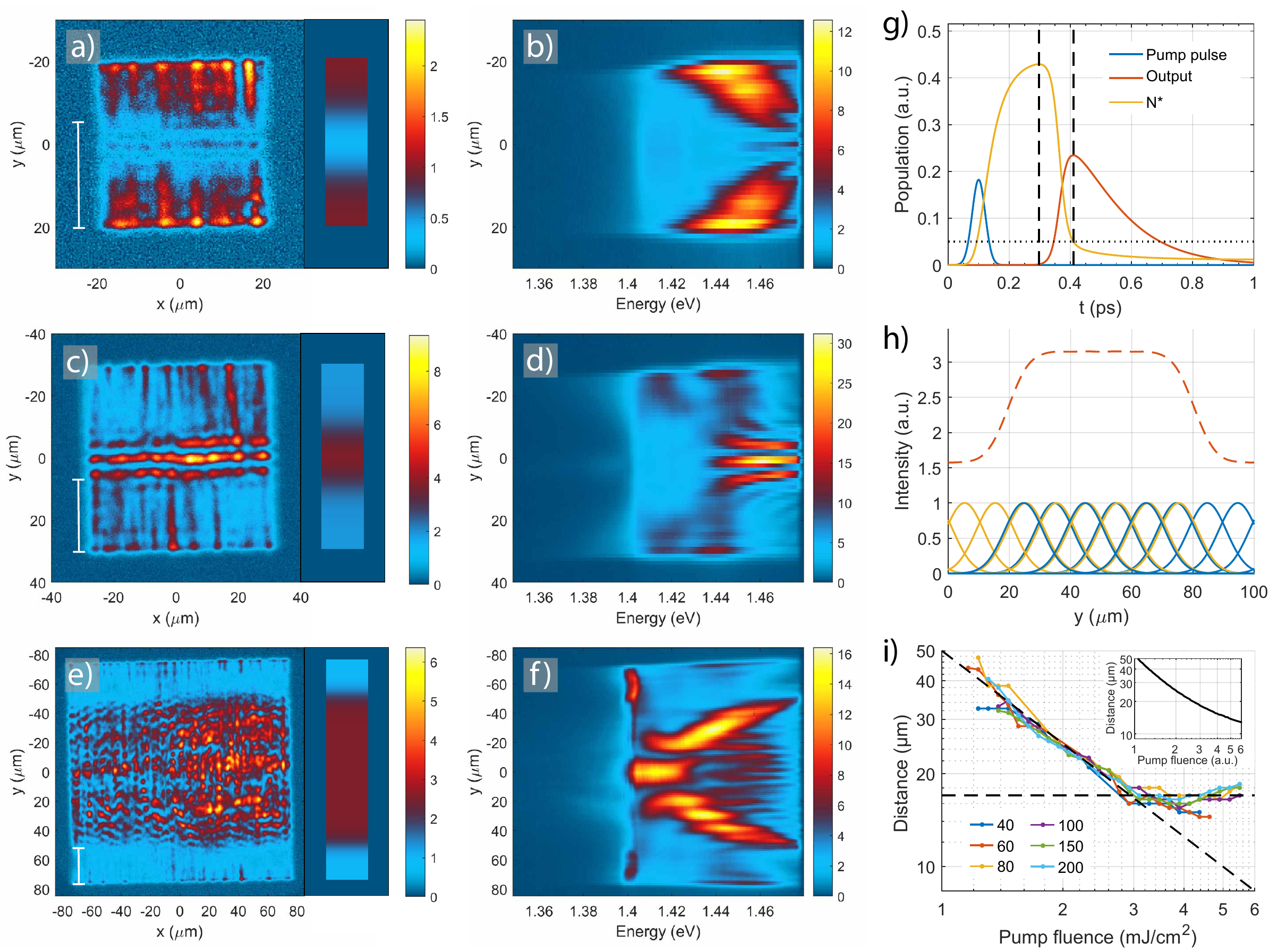}
   \caption{\textbf{Observation of stimulated emission pulse build-up in finite size lattices.}
   a)-f) Real space images (left column) and spectra (middle column) at intermediate pump fluence (2.2~mJcm$^{-2}$) for lattice sizes (a-b) 40$\times$\SI{40}{\micro m^2}, (c-d) 60$\times$\SI{60}{\micro m^2}, (e-f) 150$\times$\SI{150}{\micro m^2}. Please note the different color scales in different panels. Line spectra for the different lattice sizes are presented in Supplementary Figure~11. (g) Rate-equation simulation of stimulated emission. The pulse build-up time relevant for our system is marked by vertical dashed lines (the peaks of the population inversion $N^*$ and of the output pulse). Horizontal dotted line indicates the value at which $N^*$ overcomes the losses in the simulation. (h) Illustration of the sum (red dashed line) of spatial intensity profiles of the thermalizing pulses propagating to left (yellow) and right (blue) starting from everywhere along the $y$-axis of the array. The Gaussian shape approximates the increasing and decreasing intensity of the excitations as they gather gain and suffer losses under propagation. The results of such sum for different array sizes are shown as insets in (a,c,e) with the same false color as the real space images. (i) Measured distance from the array edge to the location where the red shift begins (indicated by the \SI{25}{\micro m} scale bars in (a,c,e)) as a function of pump fluence for different lattice sizes. The legend indicates the square-array sizes in \SI{}{\micro m}. The diagonal dashed line is the inverse of the pump fluence ($50/P$), and the horizontal dashed line indicates the saturation value ($\sim$\SI{18}{\micro m}). The inset shows the pulse build-up time (converted to distance by multiplying with the group velocity of the SLR mode) obtained from the rate-equation simulation. 
  }
\end{figure}

As the observed distribution does not match with either classical MB or RJ distribution, one can ask how does a distribution resembling the equilibrium Bose-Einstein (BE) case form in a non-steady-state, driven-dissipative system. In the weak coupling regime, the answer is known. Our system is similar to the photon condensates~\cite{klaers_bose-einstein_2010,kirton_nonequilibrium_2013,schmitt_thermalization_2015} in the sense that dye molecules with a vibrational level structure provide the thermalization mechanism. Differences to the photon condensates are our type of excitations (plasmonic-photonic lattice modes), ultrafast time-scales, and strong light-matter coupling. It has been shown both experimentally and theoretically that, in the weak coupling regime, recurrent absorption and emission processes of light with molecules whose vibrational manifold is coupled to an external bath lead to thermalization and condensation with a BE(-type) distribution. This occurs both for continuous wave~\cite{klaers_bose-einstein_2010,kirton_nonequilibrium_2013,schmitt_thermalization_2015,weill_bose-einstein_2019} and pulsed~\cite{hakala_bose-einstein_2018,Greveling2018,walker2019nonstationary} pumping.
The vibrational manifold serves as the energy loss channel to move the photon population towards lower energies, and thermal population of the vibrational states provides the temperature for the BE distribution. Due to the vibrational energy loss, the molecules may emit at lower energy than they absorb: this provides an effective coupling between photons at different energies, thus the thermalization process does not need scattering originating from Coulomb interaction as in inorganic semiconductor polariton condensation. The thermalization requires that several~\cite{KirtonPRA2015,chiocchetta_laser_2017} (or just one~\cite{schmitt_dynamics_2018}) absorption-emission cycles take place within the lifetime of the system. The speed of the thermalization process in general depends on the number of molecules, strength of the light-matter coupling, and the number of photons since stimulated processes may be important. It is plausible that this mechanism or its modified form provides the thermalization also in our present strongly coupled case. Although the lifetime of our system is extremely short, the emission-absorption processes are, as we show, highly stimulated during the red shift, also at the higher energies. This helps to fulfill the thermalization criterion of several emission-absorption cycles within the lifetime. The thermalization rates in the present case are higher (e.g., 0.20 eV/ps in Figure~2e) than observed for the same molecular concentration in our previous work~\cite{hakala_bose-einstein_2018} (0.08 eV/ps), further confirming that the larger number of photons and stimulated processes are contributing to the speedup.

A discrete step of simultaneous creation of a low energy polariton and a vibrational quantum is often quoted as the route for organic semiconductor polariton condensation~\cite{kena-cohen_room-temperature_2010,ramezani_plasmon-exciton-polariton_2017,somaschi_ultrafast_2011,yamashita_ultrafast_2018,Ramezani2018}. The microscopic foundation of this phenomenon is the same as that of photon condensation, but the parameter regimes differ. A discrete step is likely to be the adequate picture when the absorption and emission spectra of the molecules show distinct vibrational sub-peaks; in contrast, a smooth red shift process leading to a BE distribution is more likely for molecules whose vibrational states are not prominently visible in the spectra~\cite{KirtonPRA2015}. Our system which shows strong coupling (polaritons), and molecule spectra with no vibrational shoulders (Figure~1d), is in the middle ground between the weak coupling photon condensation mechanism and the relaxation by a discrete step, and requires a theoretical description going beyond the approximations done in both. For a single molecule with one vibrational state coupled to a few light modes, one can theoretically describe how a process that is coherent, except for vibrational losses that provide the energy loss channel, leads to rapid red shift of emission (Supplementary Figure~5 and Note~6). To rigorously describe our observations, one would need a model consisting of many molecules with several vibrational states coupled to a thermal reservoir, and multiple light modes at the multi-photon regime. The model should then be solved without resorting to weak coupling perturbation theory in the light-matter coupling. The current state-of-the-art theory~\cite{herrera_theory_2018,strashko_organic_2018,del_pino_tensor_2018,shammah_open_2018,radonjic_interplay_2018} forms a good starting point for this kind of advanced description. Such theory could predict how the condensation threshold depends on losses, thermalization speed, and competition with lasing, which are expected to play a role since the photon numbers emitted by the condensate that we observe here are several orders of magnitude larger than the equilibrium estimate for the critical number~\cite{hakala_bose-einstein_2018}.

\subsection*{Effect of pulse duration}

The spatial measurements have enabled an astounding, yet indirect, way to look into the dynamics of the system. To complement the spatial observations, we have probed the dynamics directly in the time domain by altering the excitation pulse duration. A 50~fs excitation pulse results in the double threshold behaviour with a distinct regime for lasing, an intermediate regime showing an incomplete stimulated thermalization process, and condensation, as explained above. Remarkably by using a longer pulse, we observe only the first (lasing) threshold, and the system does not undergo condensation even at higher pump fluences. The real space intensity distribution and spectrum remain almost unchanged from low to high pump fluence for a 500~fs pulse duration, Figure~6. The intensity distributions and spectra resemble the lasing regime seen at low pump fluence with the 50~fs pulse (Figure~2b-c), while the intermediate and condensation regimes are absent. Besides the luminescence intensity, the different threshold behaviour is clearly visible in the FWHM curves of the spectral maximum (Supplementary Figure~6 and Note~7). The FWHM is significantly decreased with both pulse durations at the first (lasing) threshold but only with the 50~fs pulse, the FWHM is decreased even further at the second (condensation) threshold. The $k$-space images and spectra for the 500~fs pulse (Supplementary Figure~7) reveal that the luminescence from low to high pump fluences is spread widely in the TM mode, hence there is no 2D confinement.

So far we have compared the results for 50~fs and 500~fs excitation pulses at the same pump fluences so that the total amount of energy injected to the system per pulse is the same for different pulse durations. However, triggering the stimulated thermalization process might depend on the instantaneous pump intensity rather than the fluence, which means that the condensation threshold could be reached also with longer pulses if the pump fluence was increased. To test this, we have further studied the dependence on pulse duration by several intermediate measurements (Figure~7) that show condensation with 100~fs and 250~fs pulses, but not with longer pulses. The condensation threshold for the 100~fs pulse is equal to that of the 50~fs pulse (3.5~mJcm$^{-2}$), whereas for the 250~fs pulse it is higher (4.5~mJcm$^{-2}$). The resulting line spectra at the condensation threshold for the 100~fs and 250~fs pulse durations (blue and red solid lines in Figure 7a) are similar to that measured for the 50~fs pulse presented in Figure 2a: macroscopic population at the band edge followed by a linear distribution at the higher energies. Fit of the tail to the MB distribution gives 316$\pm$2~K and 331$\pm$4~K for the 100~fs and 250~fs pulse, respectively (see Methods for more information). With a longer pulse (350~fs), thermalization in the time-integrated signal remains incomplete (too much population in the higher energy states). With the longest excitation pulses ($>$350~fs), we observed no signs of thermalization or condensation even at the highest pump fluences that we could measure until damaging the samples. For all pulse lengths, the thermalization process competes from the same gain with the lasing. It seems that for a long excitation pulse the instantaneous population inversion does not reach a high enough value for the thermalization process to take over the lasing which is already triggered at the first threshold (see also Figure~6).

The observations for different pulse durations are summarized by Figure 7c which shows the thresholds for lasing and condensation, as well as the start of the incomplete stimulated thermalization regime, as a function of pulse duration. The dependence of the condensation and incomplete stimulated thermalization thresholds on the pulse length is stronger than in the lasing case where the threshold is quite monotonous as a function of the pulse duration. Sensitivity to the excitation pulse duration highlights the ultrafast nature of plasmonic systems and endorses the sub-picosecond dynamics of the thermalization process.

Interestingly the critical pulse duration for observing condensation is similar to or smaller than the time ($\sim$250$-$350~fs) in which the polaritons propagate from the edges to the center in the 100$\times$\SI{100}{\micro m^2} array (see Methods for discussion on the group velocity). Besides the critical pulse duration, observing the condensation requires that the thermalization time matches the propagation time so that the polaritons have red shifted to the band edge energy while still having large population density. This is achieved by an optimal balance between the dye concentration and pump fluence (thermalization speed), lattice size (distance that the excitations need to propagate from the array edges to the centre), and lattice period (band edge energy). The condensation is also sensitive to the incident angle of the pump. To obtain a clear tail with MB distribution in the time-integrated luminescence signal, the pump pulse needs to come at normal incidence, and even a slight misalignment changes both the real space intensity pattern and the spectral distribution. Non-zero incidence angle results in a time difference of the excitation at the two edges of the array, and can cause an asymmetry in the populations of counter-propagating polariton modes.

\begin{figure}[H]
  \centering
    \includegraphics[width=1.0\textwidth]{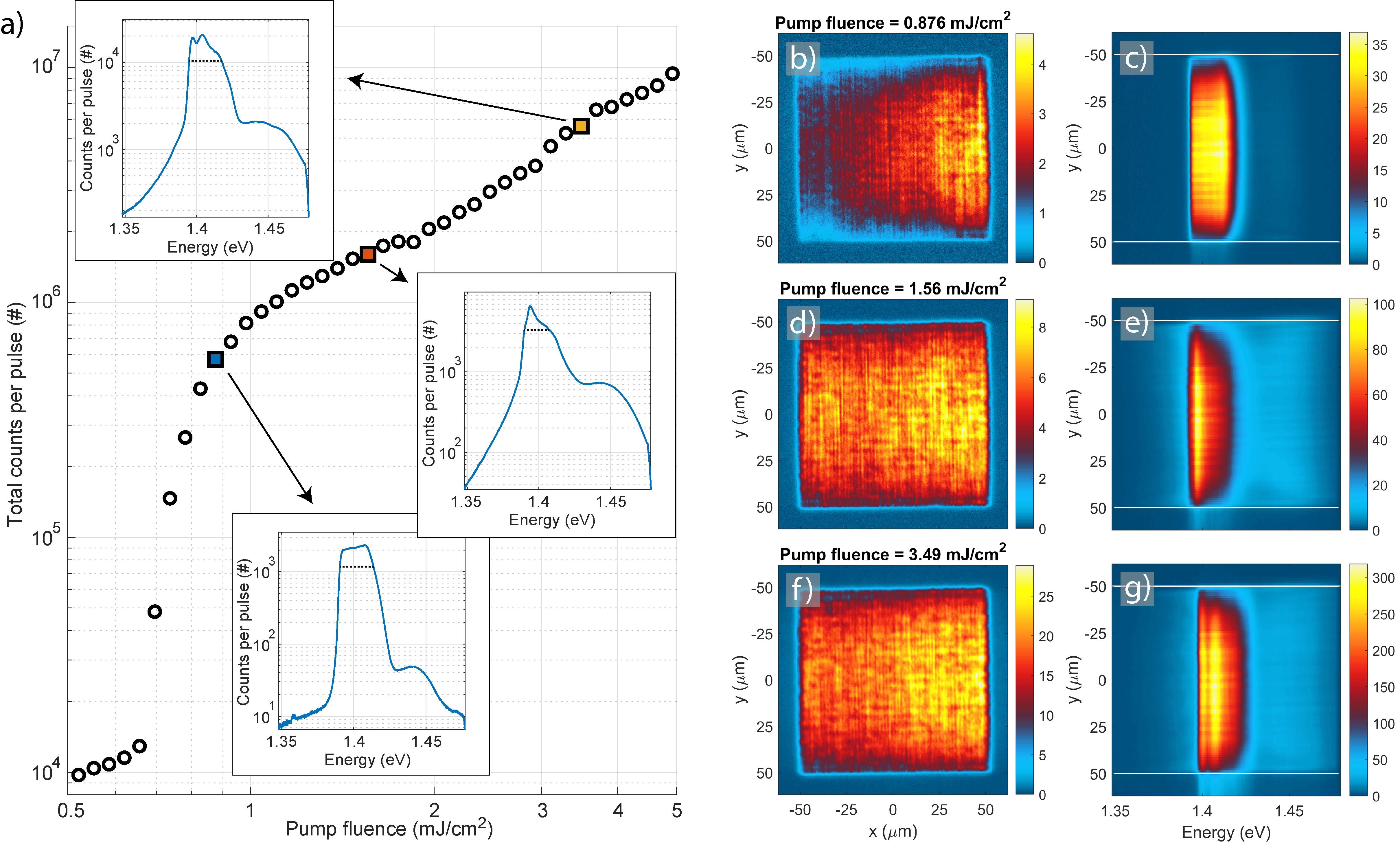}
   \caption{\textbf{Pump fluence dependence, real space images and spectra for 500~fs pump pulse duration.} (a) Pump fluence dependence of the total luminescence intensity. Insets: Line spectra obtained by integrating over the real space spectra in the $y$-direction (between the white lines). (b-g) Left column: Real space images of the plasmonic lattice. Right column: Spectral information of the luminescence as a function of position in the $y$-axis. The pump fluence is: (b-c) 0.88~mJcm$^{-2}$, (d-e) 1.6~mJcm$^{-2}$, (f-g) 3.5~mJcm$^{-2}$. FWHM of the spectral peaks is marked in the insets. With increasing pump fluence, we obtain a FWHM of 23~meV, 18~meV, and 23~meV. 
  }
\end{figure}

\begin{figure}[H]
  \centering
    \includegraphics[width=0.95\textwidth]{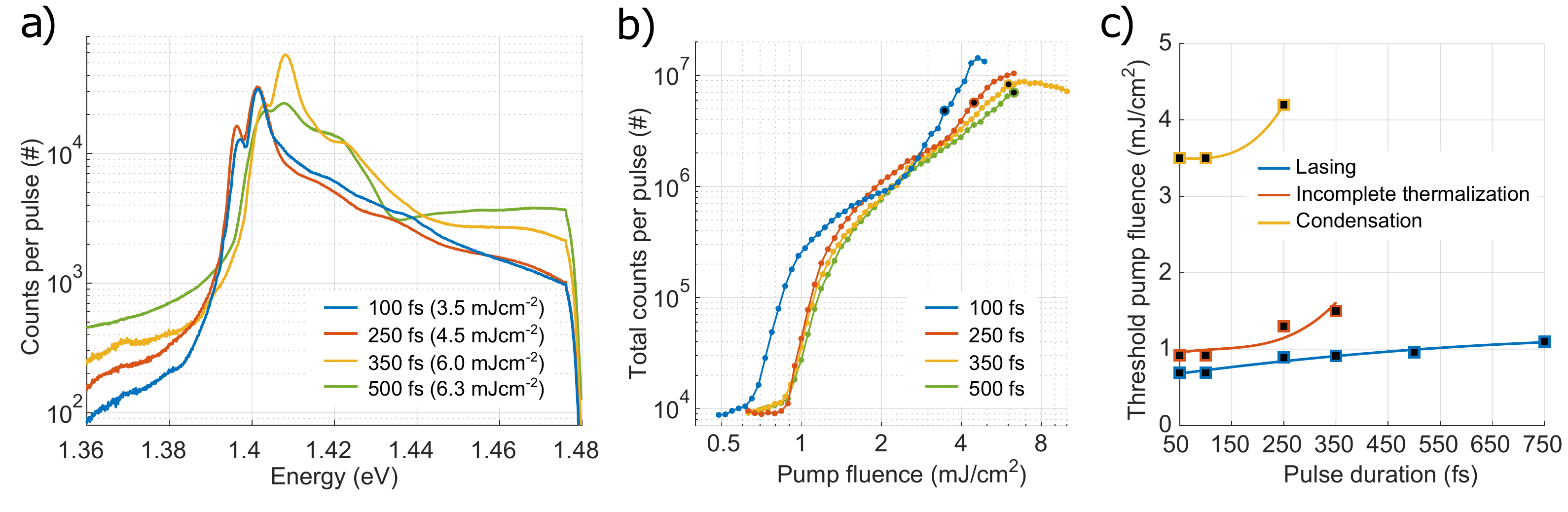}
   \caption{\textbf{ Line spectra and threshold pump fluences for different pulse durations.} (a) Line spectra for different pulse durations. Condensation occurs for 100~fs and 250~fs pulses, but not for longer ones. For the 100~fs and 250~fs pulses, the spectra are shown at the condensation threshold (3.5~mJcm$^{-2}$ and 4.5~mJcm$^{-2}$, respectively). The threshold in case of the 100~fs pump pulse is the same as for the 50~fs pulse. We have defined the condensation threshold such that a narrow peak is observed at the band edge together with a linear distribution at high energies in the time-integrated signal. For a 350~fs pulse no condensation is observed, but we show a spectrum for which the tail of the distribution is closest to linear (at 6.0~mJcm$^{-2}$). For a 500~fs pulse the spectrum is shown at the highest measured pump fluence (6.3~mJcm$^{-2}$). (b) Total luminescence intensity as a function of pump fluence. Increasing the pump pulse duration increases slightly the first (lasing) threshold, whereas the second (condensation) threshold is visible only for the 100~fs and 250~fs pulses. For the 350~fs pulse, we increased the pump fluence up to 10~mJcm$^{-2}$, after which the nanoparticle array starts to damage, to confirm that no condensation threshold occurs even at higher fluences. Rather a saturation and degradation of the luminescence signal can be observed. The highlighted data points correspond to the line spectra in (a). (c) Threshold pump fluence for lasing and condensation, as well as the beginning of the incomplete stimulated thermalization regime, as a function of pulse duration.  In the lasing regime, there is one spectral intensity maximum corresponding to the band edge (this is the lasing peak), and another one (smaller in intensity) at higher energy. First, for increasing pump fluence, the lasing peak grows and the higher energy peak diminishes. At a certain pump fluence, however, the higher energy peak starts growing. We determine this pump fluence value from the line spectra and define it as the threshold for the incomplete thermalization regime. At this pump fluence, the interference patterns (such as in Figure~2e) become visible in the real space spectra. The lines are guides to the eye.
   }
\end{figure}

\clearpage
\section*{Discussion}

Fundamental questions on the dynamics of Bose-Einstein condensation in driven-dissipative systems are still largely open, despite years of research~ \cite{carusotto_quantum_2013,byrnes_exciton-polariton_2014,altman_two-dimensional_2015,keeling_superfluidity_2017}. What is the nature of the energy relaxation and thermalization processes, how does the condensate form, and what are its quantum statistical and long-range coherence properties? These questions have been addressed for weakly coupled BECs, but become challenging to answer for strongly coupled room temperature condensates as higher energy scales imply faster dynamics. Here we have shown that plasmonic lattices offer an impressive level of access and control to the sub-picosecond dynamics of condensate formation via propagation of excitations and the finite system size. 

We experimentally demonstrated that a bosonic condensate with a clear-cut thermal excited state population can form in a timescale of a few hundreds of femtoseconds. We propose that this extraordinary speed of thermalization is possible because the process is partially coherent due to strong light-matter coupling and stimulated emission. Strong light-matter coupling at the weak excitation limit was indicated by our reflection measurements. By varying the lattice size, we revealed  the stimulated nature of the thermalization process. While strong light-matter coupling at the multi-photon regime is described by the Dicke model~\cite{kirton_introduction_2019}, to explain both the red shift and the thermal distribution we observed, one would need to involve vibrational degrees of freedom that are (strongly) coupled to the electronic ones as well as to a thermal bath. Work toward surmounting such theoretical challenges has already begun since systems where light and molecular electronic and vibrational states are strongly coupled have promise for lasing and condensation, energy transfer, and even modification of chemical reactions~\cite{ribeiro_polariton_2018,hertzog_strong_2019}. 
We have shown here that plasmonic lattices offer a powerful platform for studies of such ultrafast light-matter interaction phenomena. The shape, size, and material of the nanoparticles can be accurately tuned, as well as the lattice geometry, composition of the unit cell, and the lattice size. This provides a large, controlled parameter space vital for testing and (dis)qualifying theoretical predictions. In particular, the dynamics can be accessed in complementary ways: through conventional time-domain techniques as well as indirectly via the propagation of excitations in the lattice.

\section*{Methods}

\subsection*{Samples}
The gold nanoparticle arrays are fabricated with electron beam lithography on glass substrates where 1~nm of titanium is used as an adhesion layer (see SEM image in Supplementary Figure~8). The nominal dimensions of the plasmonic lattice, optimized for the condensation experiment, are the following: a nanoparticle diameter of 100~nm and height of 50~nm, the period in $y$- and $x$-direction of $p_y$ = 570 and $p_x$ = 620~nm, respectively, and a lattice size of \SI{100}{}$\times$\SI{100}{\micro m^2}. In reference measurements, the period $p_y$ is varied between 520 and 590~nm and the lattice size between \SI{40}{}$\times$\SI{40}{} and \SI{200}{}$\times$\SI{200}{\micro m^2}.

Asymmetric periodicity separates the diffracted orders in the energy spectrum for the two orthogonal polarizations ($e_x$ and $e_y$), and the SLR dispersions are correspondingly separated, which simplifies the measured spectra. For $x$-polarized nanoparticles (as in our experiments), the TE and TM modes correspond to combinations of ($e_x$, $k_y$) and ($e_x$, $k_x$), respectively. Under pumping, which SLR mode is mainly excited is determined by the pump polarization as the molecules are excited more efficiently via the single particle resonance at the plasmonic hot-spots of each nanoparticle~\cite{wang_band-edge_2017}.  In the experiment with different periods, $p_x$ was kept always 50~nm larger than $p_y$. In the experiment where lattice size was varied, however, the lattice period in $x$ and $y$ directions was the same ($p_x = p_y = 570$~nm). We found that asymmetric periodicity does not play a crucial role in forming the condensate but just simplifies the data analysis of the experimental results.

Group velocity for the TE mode is obtained close to the $\Gamma$-point, in the linear part of the dispersion, for samples both without and with the dye molecules. The group velocity is 0.65c for the uncoupled TE mode (Figure~1c) and 0.48c for the strongly coupled polariton mode (Figure~1d; c is the speed of light).
We use the group velocity to convert propagation distance to time. In the experiments we see that the strongly coupled dispersion starts to resemble the uncoupled one at high pump fluences due to the saturation effects, as explained in the manuscript. We cannot exactly specify what the group velocity is at a certain pump fluence, therefore we present the time conversions with a group velocity range from 0.48c to 0.65c. This means that the propagation of $50\mu m$ distance along the array takes 250$-$350~fs.

The dye molecule solution is index-matched to the glass substrate (n = 1.52), the solution being a mixture of 1:2 DMSO:Benzyl Alcohol. The solution is sealed inside a Press-to-Seal silicone isolator chamber (Sigma-Aldrich) between the glass substrate and superstrate. The dye solution has a thickness of $\sim$1~mm, which is very large compared to the extent of the SLR electric fields~\cite{hakala_lasing_2017,daskalakis_ultrafast_2018,wang_band-edge_2017}. Given by the excess of the dye molecules and the natural circulation of the fluid, there are always fresh dye molecules available for consecutive measurements of the sample when scanning the pump fluence, which makes the sample extremely robust and long-lasting. IR-792 perchlorate ($C_{42}H_{49}ClN_2O_4S$) was chosen as the dye molecule because it dissolves to the used solvent in very high concentrations, in contrast to many other dye molecules, e.g., IR-140 that has also been used by us \cite{hakala_lasing_2017,daskalakis_ultrafast_2018} and others \cite{zhou_lasing_2013,yang_real-time_2015,wang_band-edge_2017} as a gain medium in plasmonic nanoparticle array lasers. We have collected the information on tested dye molecules and clarify the reasoning behind the molecule choice in Supplementary Table 1.

\subsection*{Transmission, reflection, and luminescence measurement setup}

A schematic of our experimental setup is depicted in Supplementary Figure~9. The same setup is used for transmission, reflection and luminescence measurements with minor modifications. The spectrometer resolves the wavelength spectrum of light that goes through the entrance slit, and each pixel column in the 2D CCD camera corresponds to a free space wavelength, $\lambda_0$, and each pixel row to a $y$-position at the slit. The $y$-position further corresponds to either an angle ($k$-space) or the $y$-position at the sample (real space). The photon energy is $E = hc/\lambda_0$ and in the case of angle-resolved spectra (dispersions) the in-plane wave vector $k_{x,y} = k_0 sin(\theta_{x,y}) = 2\pi/\lambda_0 sin(\theta_{x,y})$, where $h$ is the Planck constant and $c$ the speed of light in free space. Next, the three different experiment types are explained, starting with the luminescence measurement where the sample is optically excited with an external pump laser. 
The excitation pulse (or pump pulse) is generated by Coherent Astrella ultrafast Ti:Sapphire amplifier. The pulse has a central wavelength of 800~nm, and at the laser output, a duration of $<~35$~fs with a bandwidth of 30~nm. The pump pulse is guided through a beam splitter and mirrors, and finally to the mirror $M1$ (see Supplementary Figure~9), which directs the pump pulse to the excitation path of the setup. We have a band-pass filter in the excitation path that is used in combination with a long-pass filter in the detection path to filter out the pump pulse in the measured luminescence spectra. The pump pulse is linearly polarized, and to filter only the horizontal component we have placed a linear polarizer after the band-pass filter. The pump fluence is controlled with a metal-coated continuously variable neutral-density filter wheel (ND wheel). The pump pulse is spatially cropped with an adjustable iris and the iris is imaged onto the sample with a help of lens $L1$ and the microscope objective. The inverted design enables exciting the sample at normal incidence, which is crucial for simultaneous excitation of the dye molecules over the sample. Excitation at normal incidence also prevents any asymmetry in the spatial excitation of the molecules around the nanoparticles with respect to lattice plane. The inverted pumping scheme and accurate optical alignment were essential for repeatable and precise condensate formation. 

In the detection path, we have the long-pass filter and optionally a linear polarizer. An iris or pinhole acts as a spatial filter at the 1st image plane to restrict the imaged area at the sample. The 1st image plane is relayed onto the real-space camera (1st Cam.). In the $k$-space measurements, the back-focal plane of the objective (Fourier plane; containing the angular information of the collected light) is relayed onto the 2D $k$-space camera (2nd Cam.) as well as onto the spectrometer slit, with the tube lens and a $k$-space lens. For the real space measurements, the beam-splitter before the $k$-space lens is replaced with an additional real-space lens to produce the 2nd image plane of the sample to 2nd Cam. and onto the spectrometer slit. The spectrometer slit selects a vertical slice either of the 2D $k$-space image or the real space image. In the luminescence measurements, we use a slit width of \SI{500}{\micro m}. In the $k$-space, it corresponds to $\pm1.3^{\circ}$ around $\theta_x=0$, or to $\pm$\SI{0.16}{\micro m^{-1}} around $k_x=0$ at $E = 1.4$~eV. Respectively in the real space, the slit opening of \SI{500}{\micro m} corresponds to \SI{27}{\micro m} slice at the sample.

The dispersion of optical modes in the bare plasmonic lattice can be measured in transmission mode of the setup, where the sample is illuminated with a focused and diffused white light from a halogen lamp. The lattice modes are visible as transmission minima (extinction maxima) in the angle-resolved spectrum. When a thick layer of high-concentration dye molecule solution is added in the chamber on top of the nanoparticle array, the transmission measurement is not applicable due to a complete absorption by the molecules. To access the dispersion of the lattice modes in this case, we use the setup in reflection mode by utilizing the same inverted design as in the luminescence measurement. The halogen lamp is inserted before the iris in the excitation path, that is imaged onto the sample, and the dispersion of the lattice modes is revealed by reflection (scattering) maxima in the collected angle-resolved spectrum. 

The luminescence measurement as a function of pump fluence is automated with LabView. Predefined fluence steps are measured such that for each step: 1) the calibrated ND filter wheel is set to a correct position, 2) the shutter is opened, 3) the image is acquired with spectrometer and optionally with 1st and 2nd Cam., and 4) the shutter is closed. Integration time of the spectrometer is automatically adjusted during the measurement to avoid saturation at highly non-linear threshold regimes. The pump pulse duration is measured with a commercial autocorrelator (APE pulseCheck 50). In the pulse duration measurement, the pump pulse goes through the same optics as in the actual experiments. The pulse duration is changed by adjusting the stretcher-compressor of the external pump laser.

\subsection*{Fits to the Maxwell--Boltzmann distribution}

We fit the thermalized tail in the measured population distributions to Maxwell--Boltzmann distribution (Figure~2 and Supplementary Figure~6). The fit function is given by
\begin{align}
f_{\mathrm{MB}}(E) = \frac{d(E)}{\mathrm{e}^{(E-\mu)/(k_\mathrm{B} T)}},
\end{align}
where $d(E)$ is the degeneracy of the modes as a function of energy $E$, $\mu$ is the chemical potential, $k_\mathrm{B}$ is the Boltzmann constant, and $T$ is temperature. The fit was done for the part of the distribution that is linear in logarithmic scale, at pump fluence at/above the threshold. We call this part of the distribution (between energies 1.41...1.47eV) the "tail".  Fitting was performed with a non-linear least squares method. 

The degeneracy $d(E)$ was approximated by the density of states for light travelling in a 2D plane. The light dispersion in the $xy$ plane forms a conical surface ($E = \hbar c /n  \sqrt{k_x^2+k_y^2}$). The dispersions of the SLR modes are well approximated by this everywhere except the very near vicinity of the $\mathbf{k}=0$ point~\cite{Moerland2017}, and our fitted range starts from a finite $\mathbf{k}$ where the approximation is valid. This dispersion results in a linearly increasing but nearly constant density of states for the fitted energy range of $\sim$60~meV ($d(E)=1...1.05$).

The fit gives a temperature of 313$\pm$2~K for a chosen pump fluence of $P = 3.5$~mJcm$^{-2}$, error limits representing the 95\% confidence bounds for the fit. This fit is presented in manuscript Figure~2a (top inset) and Supplementary Figure~6a. The fitted pump fluence is chosen such that the fluence is the lowest showing a linear slope in the time-integrated population distribution (in logarithmic scale). The chosen fluence also corresponds to the narrowest FWHM of the highest condensate peak (see Supplementary Figure~6c). 
The high-energy tail remains linear over pump fluences between $\sim$3.5...4~mJcm$^{-2}$, with a slightly changing slope. For two higher pump fluences of 3.7 and 3.9~mJcm$^{-2}$, the fits to the the Maxwell--Boltzmann distribution give temperatures of 282$\pm$2~K and 250$\pm$2~K, respectively. Goodness of fit is described by the square root of the variance of the residuals (RMSE) and the R-square value. The values of (RMSE, R-square) for the three pump levels 3.5, 3.7, and 3.9~mJcm$^{-2}$ are ($107, 0.996$), ($98, 0.998$), and ($146, 0.998$), respectively. For pump fluences above $\sim$4~mJcm$^{-2}$, the linear slope is distorted and the condensate degrades. This is evident also as a decrease of the spatial coherence (Figure~4a) and an increase of the FWHM of the spectral maximum (Supplementary Figure~6c).

For the longer pulses, 100~fs and 250~fs, the fit gives 316$\pm$2~K and 331$\pm$4~K at the condensation threshold, 3.5~mJcm$^{-2}$ and 4.5~mJcm$^{-2}$, respectively. The values of (RMSE, R-square) for the fits are ($119, 0.997$) and ($219, 0.985$), respectively. The fits are still quite good for these pulse durations, whereas for 350~fs and longer pulses no thermal Maxwell–Boltzmann distribution was observed.

\subsection*{Estimation of photon number in the condensate}

The photon number in the condensate is estimated from the measured luminescence intensity. A strongly attenuated beam from the external pump laser (800~nm, 1~kHz) is directed to the spectrometer slit, and the total counts given by the spectrometer CCD camera (Princeton Instruments PIXIS~400F) is compared to the average power measured with a power meter (Ophir Vega). The measured average power of 167~nW corresponds to $6.7 \cdot 10^8$ photons/pulse whereas the total counts in the CDD camera are $8.4 \cdot 10^6$, leading to a conversion factor of $\sim$80 photons/count. 
In the condensation regime, the total counts per pulse is about $3 \cdot 10^6$ (manuscript Figure~2a). The collection optics including the beam splitters reduce the signal roughly by a factor of 2.5, and as the slit width of \SI{500}{\micro m} corresponds to \SI{27}{\micro m} at the sample, we collect luminescence from an area that is about 1/4 of the \SI{100}{\micro m} wide nanoparticle array. Finally, the sample is assumed to radiate equally to both sides, so the actual photon number per emitted condensate pulse becomes: $n_{ph} \approx 2.5 \times 4 \times 2 \times 80 \times 3 \cdot 10^6 = 4.8 \cdot 10^9$. 

\subsection*{Spatial coherence measurements}

Spatial coherence of the sample luminescence is measured with a Michelson interferometer. The real space image is split into two arms and the image in one of the arms is inverted in vertical direction with a hollow roof retro-reflector. Then the two images are combined again with a beam splitter and overlapped at the camera pixel array, simultaneously. With this design the spatial coherence (first-order correlation $g^{(1)}$) can be measured separately along both $x$- and $y$-axis of the plasmonic lattice. 
The retro-reflector always inverts the image vertically, with respect to $y=0$ in laboratory reference frame, but the sample and the pump polarization can be rotated 90$^\circ$ to measure $g^{(1)}(\boldsymbol{-x},\boldsymbol{x})$ instead of $g^{(1)}(\boldsymbol{-y},\boldsymbol{y})$ in the lattice coordinates. 
The first-order correlation function describing the degree of spatial coherence is given by 
\begin{align}
g^{(1)}(\boldsymbol{-y}, \boldsymbol{y}) = \frac{\langle E^*(\boldsymbol{-y}) E(\boldsymbol{y}) \rangle}{\sqrt{\langle E(\boldsymbol{-y})^2 \rangle \langle E(\boldsymbol{y})^2 \rangle}},
\end{align}
where $E(\boldsymbol{y})$ is the electric field at point $\boldsymbol{y}$. The first-order correlation function relates to the interference fringe contrast $C$ as follows: 
\begin{align}
\label{contrast}
C(\boldsymbol{y}, -\boldsymbol{y}) = \frac{2~ \sqrt[]{I(\boldsymbol{y})I(-\boldsymbol{y})}}{I(\boldsymbol{y})+I(-\boldsymbol{y})} g^{(1)}(\boldsymbol{y}, -\boldsymbol{y}),
\end{align}
where $I(\boldsymbol{y})$ is the luminescence intensity at point $\boldsymbol{y}$ of the lattice. The fringe contrast in the interfered images is extracted with a Fourier analysis as explained in Supplementary Note~4.

\section*{Data availability}
The data that support the findings of this study are available in zenodo.org with the identifier DOI: 10.5281/zenodo.3648650~\cite{our_bec_data}.

\clearpage
\section*{Acknowledgments}
We thank Janne Askola for his help in intensity calibration of the spectrometer. We thank Jonathan Keeling and Kristín Arnardóttir for useful discussions. \textbf{Funding:} This work was supported by the Academy of Finland under project numbers 303351, 307419, 327293, 318987 (QuantERA project RouTe), 322002, 320166, and 318937
(PROFI), by Centre for Quantum Engineering (CQE) at Aalto University, and by the European Research Council (ERC-2013-AdG-340748-CODE). This article is based on work from COST Action MP1403 Nanoscale Quantum Optics, supported by COST (European Cooperation in Science and Technology). Part of the research was performed at the Micronova Nanofabrication Centre, supported by Aalto University. The Triton cluster at Aalto University (Science-IT) was used for computations. A.J.M acknowledges financial support by the Jenny and Antti Wihuri Foundation. K.S.D. acknowledges financial support by a Marie Skłodowska-Curie Action (H2020-MSCA-IF-2016, project id 745115). \textbf{Author contributions}: P.T. initiated and supervised the research. A.I.V. fabricated the samples. A.I.V., A.J.M., and K.S.D. conducted the experiments. A.J.M., A.I.V., P.T., and T.K.H. did the data analysis. A.I.V. performed the rate-equation, A.J.M. the quantum model, and M.N. the T-matrix calculations. A.I.V., A.J.M., and P.T. wrote the manuscript with all authors. \textbf{Competing interests:} The authors declare no competing financial interests.

\clearpage

\setcounter{figure}{0} 
\setcounter{table}{0}
\makeatletter
\renewcommand{\figurename}{Supplementary Figure}
\renewcommand{\tablename}{Supplementary Table}
\makeatother

\section*{Supplementary Information}

\begin{figure}[ht!]
  \centering
    \includegraphics[width=1.0\textwidth]{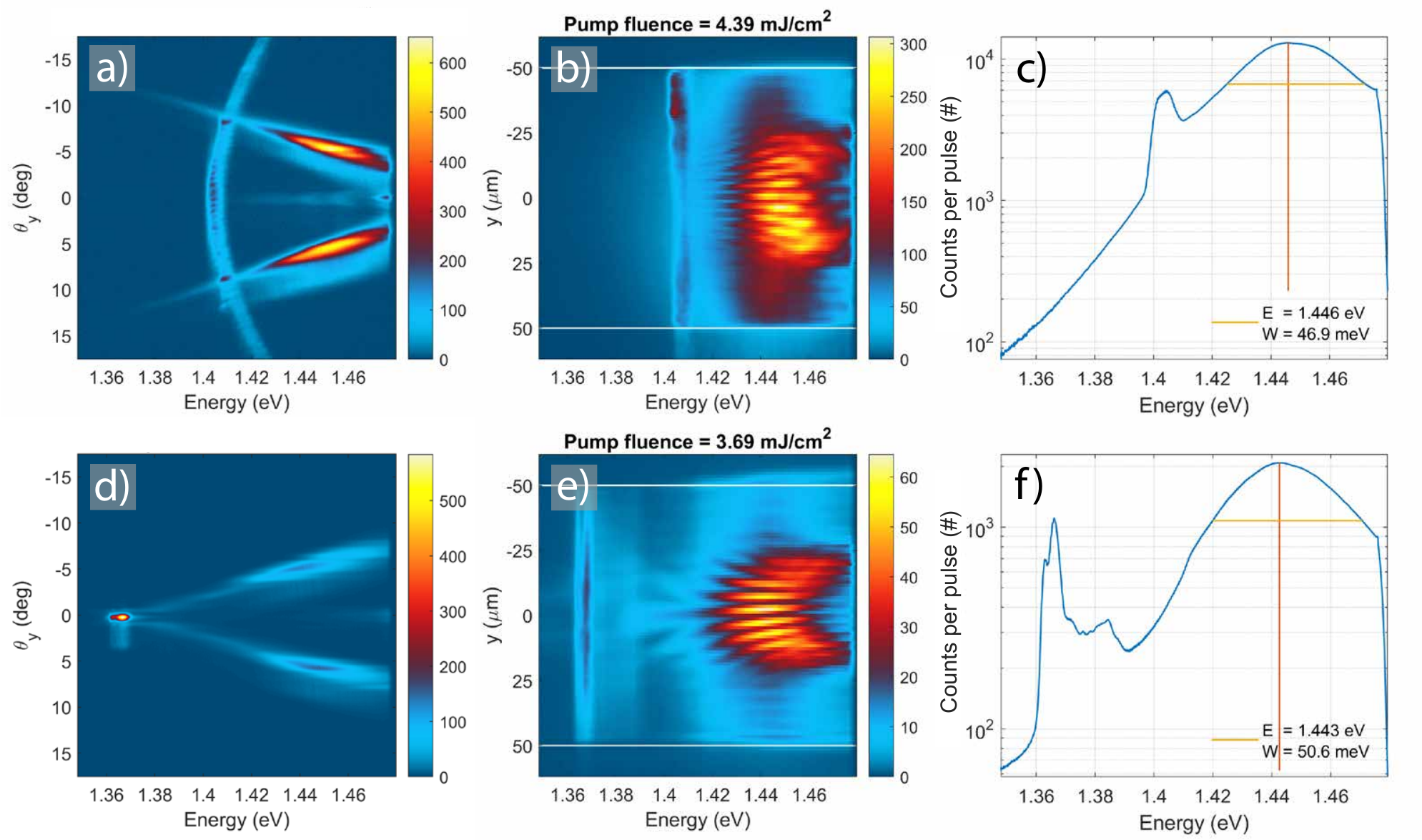}
   \caption{\textbf{Effect of the band-edge energy.} Left column: $k$-space spectra in the TE mode direction. Middle column: Real space spectra. Right column: Line spectra obtained by integrating over the real space spectra in $y$-direction (between white lines). Results are shown for two lattice periods: (a-c) $p_y = 520$~nm and (d-f) $p_y = 590$~nm. Note that in (a) the curved band visible at 1.40...1.42~eV is the TM mode of the $y$-polarized nanoparticles ($p_x = 570$~nm). 
  }
\end{figure}

\begin{figure}[ht!]
  \centering
    \includegraphics[width=0.5\textwidth]{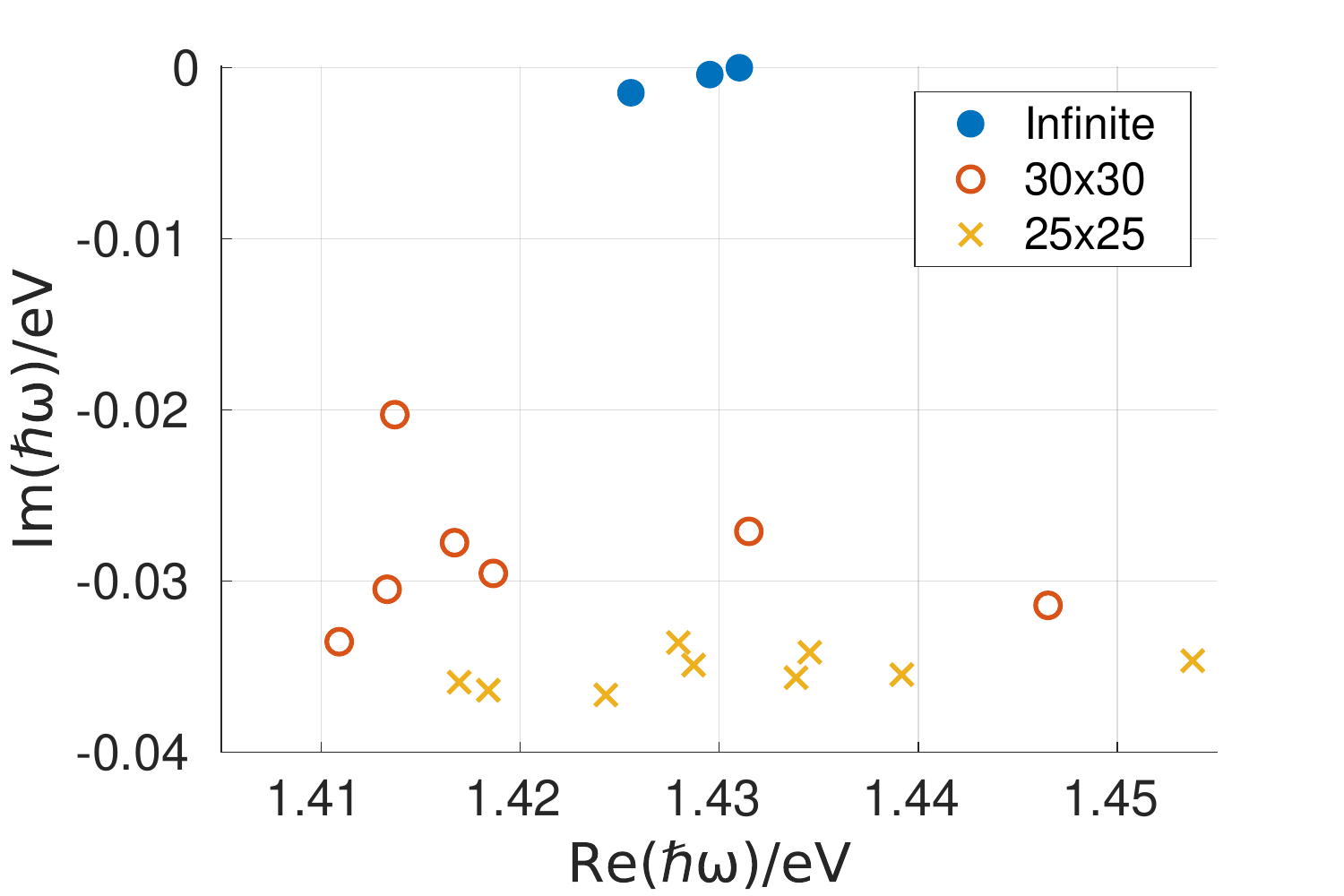}
   \caption{\textbf{Results of the T-matrix simulation for infinite and finite arrays of nanoparticles.} Modes at $k=0$ of infinite and finite rectangular arrays of cylindrical nanoparticles with periods $p_x=\SI{621}{nm}$, $p_y=\SI{571}{nm}$, radii $R=50$~nm and height $h=50$~nm, obtained by the multiple-scattering $T$-matrix method up to the octupole degree. Real and imaginary (representing loss) parts of the three distinct modes of an infinite lattice are shown by the blue dots. Mode energies (real and imaginary parts) for finite lattices of with 20$\times$20 particles (yellow crosses) and 30$\times$30 nanoparticles (red circles) show larger number of modes and also higher losses.}
   \label{fig:TMatrixInfSpherical}
\end{figure}

\begin{figure}[ht!]
  \centering
    \includegraphics[width=1.0\textwidth]{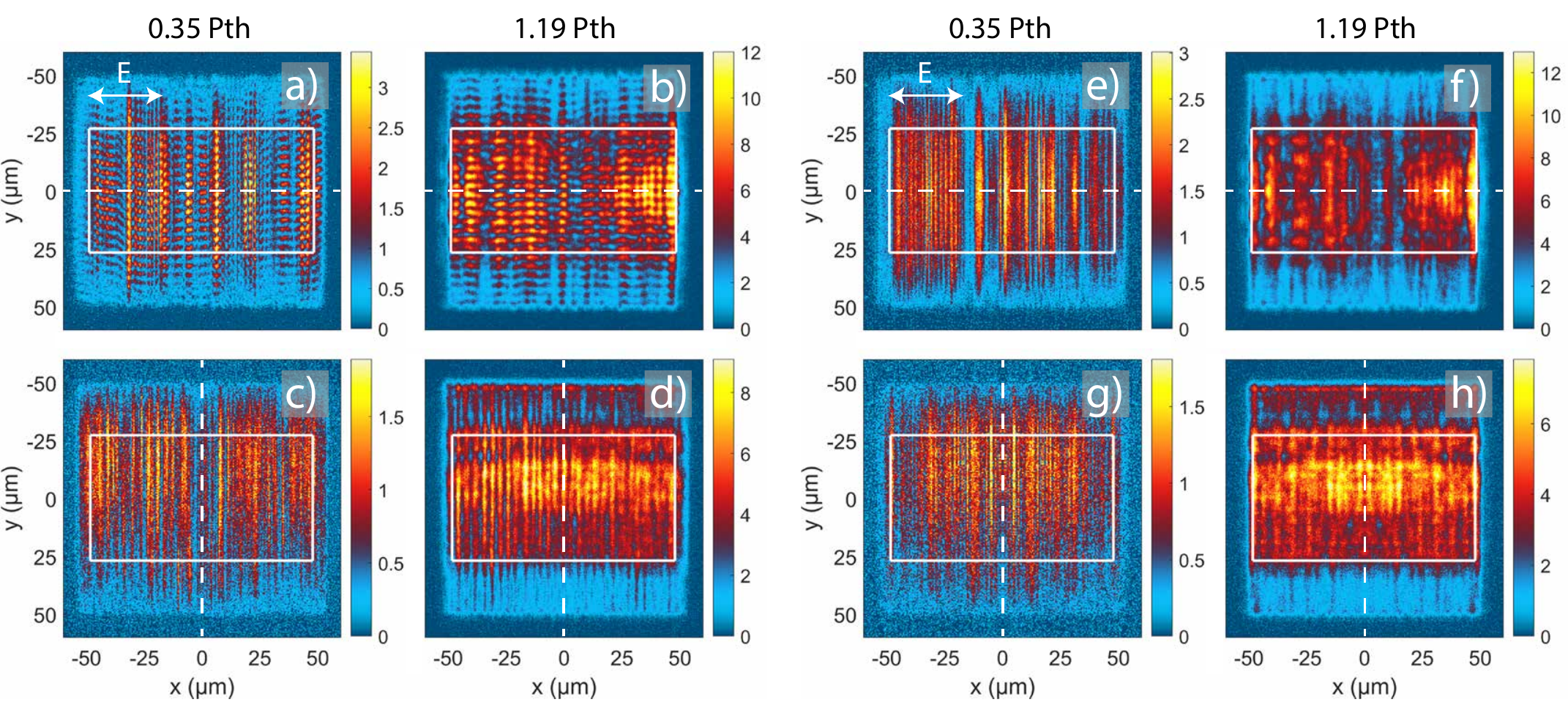}
   \caption{\textbf{Spatial coherence measurement along $y$ and $x$ axis of the plasmonic lattice, with incoherently summed reference images.} (a-d) show the same images as the main text Figure~4, (e-h) show reference images for the same pump fluences. The reference images are obtained by inverting a real space image from one of the Michelson arms in the post processing, and summing it with the original non-inverted image. This is an incoherent equivalent of the coherently summed images in the Michelson interferometer. There is a clear difference between the interfered and non-interfered images in the $y$-direction coherence measurement (a,b,e,f) but not so clear in the $x$-direction measurement (c,d,g,h). For that, Fourier analysis of spatial frequencies is necessary to extracting the fringe contrast caused by spatial coherence. Here $P_{th}$ means the threshold pump fluence for condensation and \textbf{E} refers to the pump polarization. The real space images are recorded for single pump pulses.
  }
\end{figure}

\begin{figure}[ht!]
  \centering
    \includegraphics[width=0.75\textwidth]{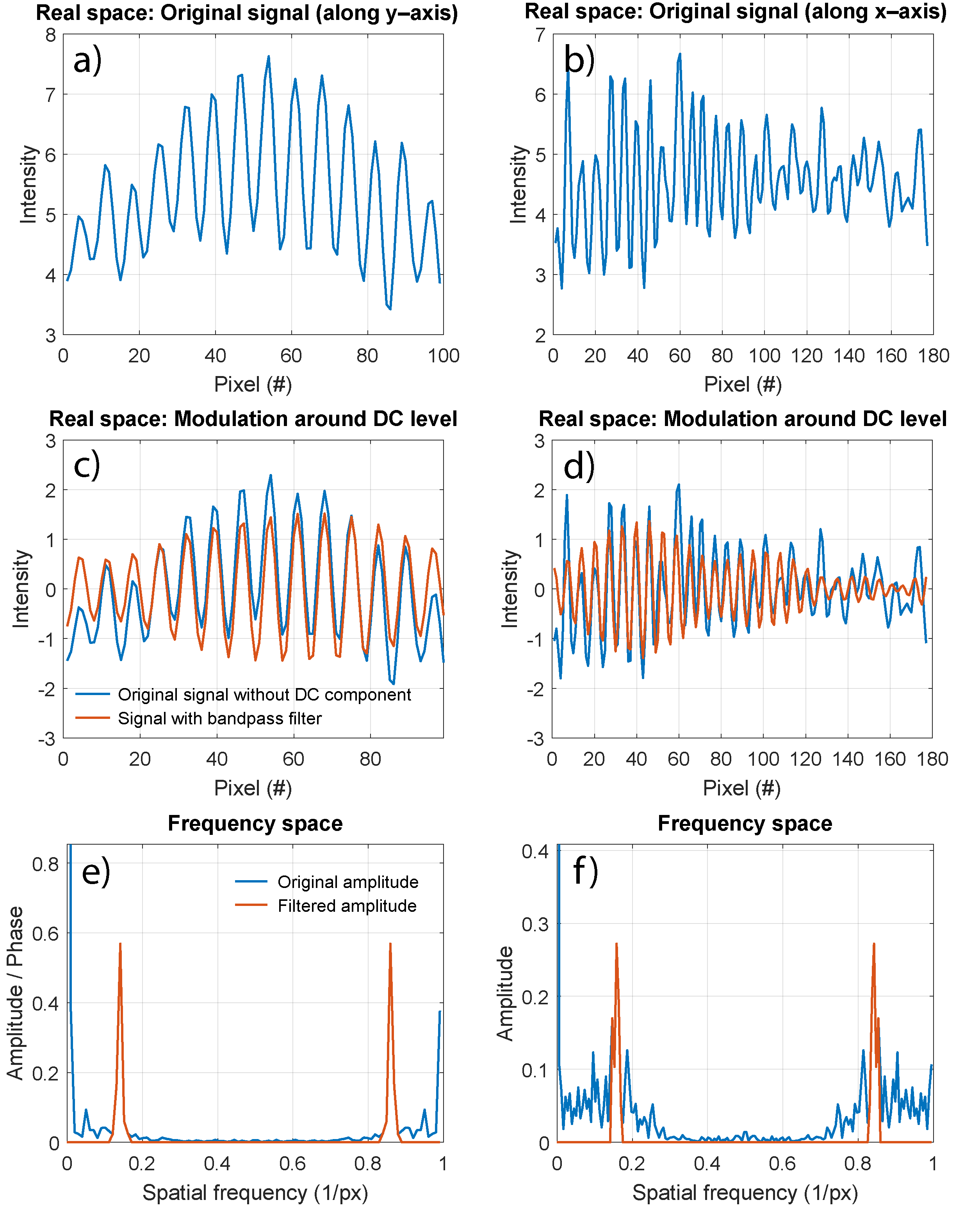}
   \caption{\textbf{Fourier analysis method for extracting the fringe contrast in the Michelson interferometer images. } The left panel (a,c,e) shows the intensity along $y$-axis of the lattice, averaged over $x$ (see white box in Supplementary Figure~3b), and the right panel (b,d,f) shows the intensity along $x$-axis of the lattice, averaged over $y$ (Supplementary Figure~3d). In (c-f), the blue curves show the original signal and the red curves the filtered signal.
  }
\end{figure}

\begin{figure}[ht!]
  \centering
    \includegraphics[width=1.0\textwidth]{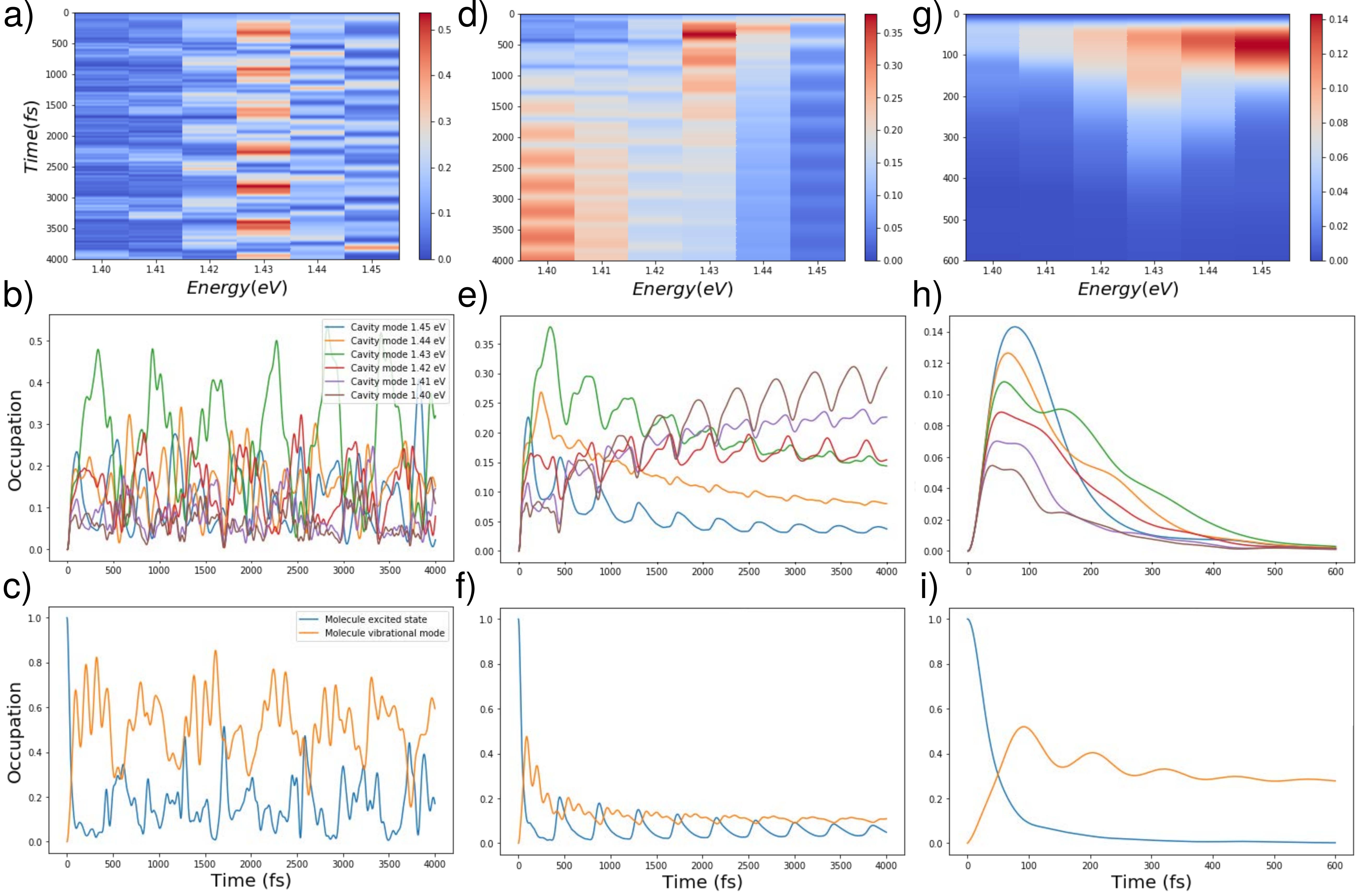}
   \caption{\textbf{Results of the quantum model.}  Time evolution of the occupation probability of cavity modes and the molecule excited state and the vibrational mode. Results are shown for the cases of (a-c) no losses, (d-f) only the vibrational mode loss, and (g-i) all losses present. The parameters used in the simulation are listed in Supplementary Note~6. Here (a,d,g) show the occupations of cavity modes at different energies in a color scale, and (b,e,h) their time evolution. Molecule electronic excited state and vibrational state population time evolution is shown in (c,f,i). When there are no losses (a-c), the occupation oscillates reversibly between different cavity modes and the molecule excited state. Applying the vibrational mode dissipation (d-f) results in red shift of the occupation towards the lowest-energy cavity mode. (g-i) shows that even in the presence of losses, the occupation reaches a lower-energy cavity mode before vanishing. Note that the cavity mode at 1.43~eV (green solid line in (b,e,h)) is favoured because it resides at the energy $\omega_m-\omega_v+\Delta$, in other words, it corresponds to a resonance where original excited state energy of the molecule is distributed to a vibrational excitation and the cavity mode. Here $\Delta$ is energy offset due to coupling between the molecule and the cavity modes. The shift can be approximated as $\Delta \approx \sum_i |g|^2/(\omega_m - \omega_i)$.
  }
\end{figure}

\begin{figure}[ht!]
  \centering
    \includegraphics[width=1.0\textwidth]{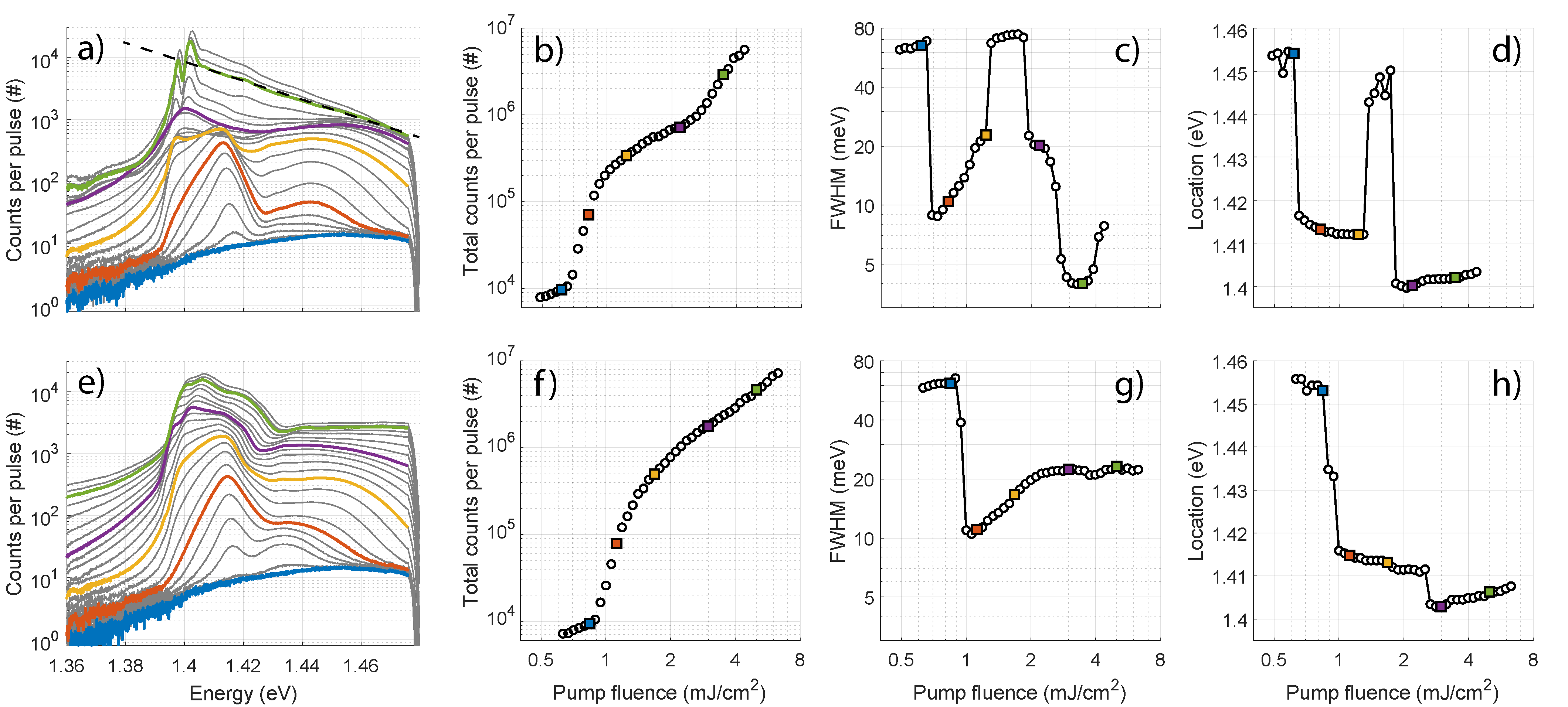}
   \caption{\textbf{Pump fluence dependence and line spectra for 50~fs and 500~fs pulse durations.} First column: Population distribution at different pump fluences. Second column: Fluence dependence of the total luminescence, summed counts under the curves in (a,e), showing the threshold behaviour. Third column: The full-width at half-maximum (FWHM) of the spectral maximum as a function of pump fluence. Fourth column: The energy position of the spectral maximum. 
   The results are shown for two excitation pulse durations: (a-d) 50~fs and (e-h) 500~fs. The pump fluences indicated by colored markers in (b-d,f-h) correspond to the colored lines in (a,e). The short excitation pulse results in the double threshold behaviour with distinct regimes of lasing, incomplete thermalization, and condensation, as explained in the main text. Before the onset of the first threshold (blue), the distribution reflects the spontaneous emission profile of the molecule. When the first threshold is reached (red), lasing peak is visible at around the band edge energy. After the first threshold (yellow and purple), population at broad range of higher energies is increased due to incomplete thermalization (intermediate regime). 
   Condensation takes place at the second threshold (green), only for the 50~fs pulse, where narrow peaks are observed at the band edge, followed by a thermalized tail at the higher energies. Fit to the Maxwell--Boltzmann distribution (dashed line in (a)) gives the temperature of 313$\pm$2~K.
  }
\end{figure}

\begin{figure}[ht!]
  \centering
    \includegraphics[width=0.85\textwidth]{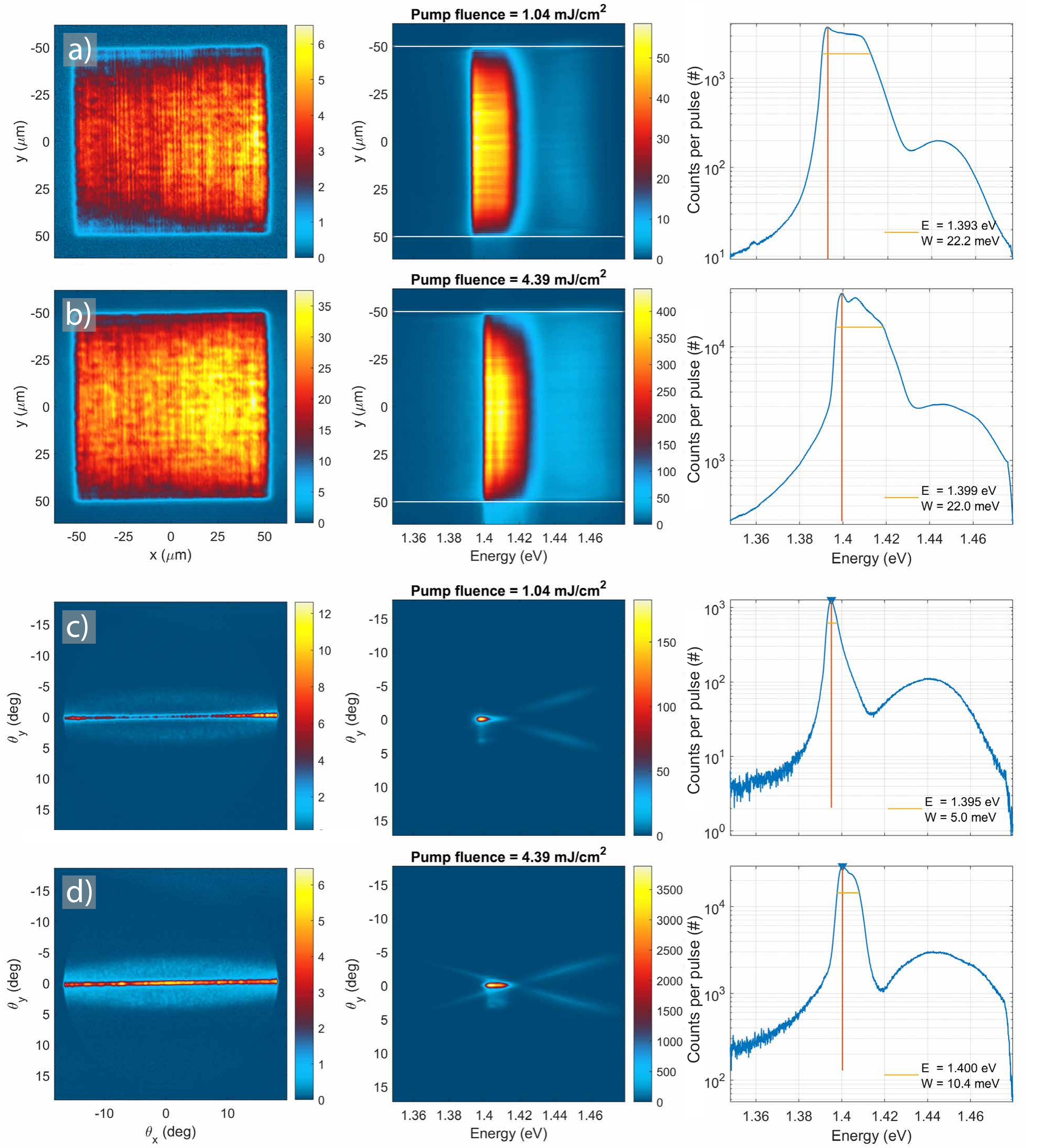}
   \caption{\textbf{Real space and $k$-space measurement with 500~fs pulse duration.} (a-b) Real space measurements. Left column: Real space images. Middle column: Real space spectra. Right column: Line spectra of the intensity integrated over the real space spectra along $y$-axis between the white lines. (c-d) $k$-space measurements. Left column: 2D $k$-space images. Middle column: $k$-space spectra. Right column: Line spectra of the intensity integrated over the $k$-space spectra. The results are shown for a low (a,c) and a high (b,d) pump fluence. With a 500~fs pulse, we observe only the lasing regime where a narrow peak occurs at the band edge and some amplified spontaneous emission occurs in the dispersion branches, both above and below the crossing point. The luminescence from low to high pump fluences is spread widely in the TM mode, with little intensity seen at higher energies in the TE mode -- no sign of thermalizing population nor 2D confinement. 
  }
\end{figure}

\begin{figure}[ht!]
  \centering
    \includegraphics[width=0.6\textwidth]{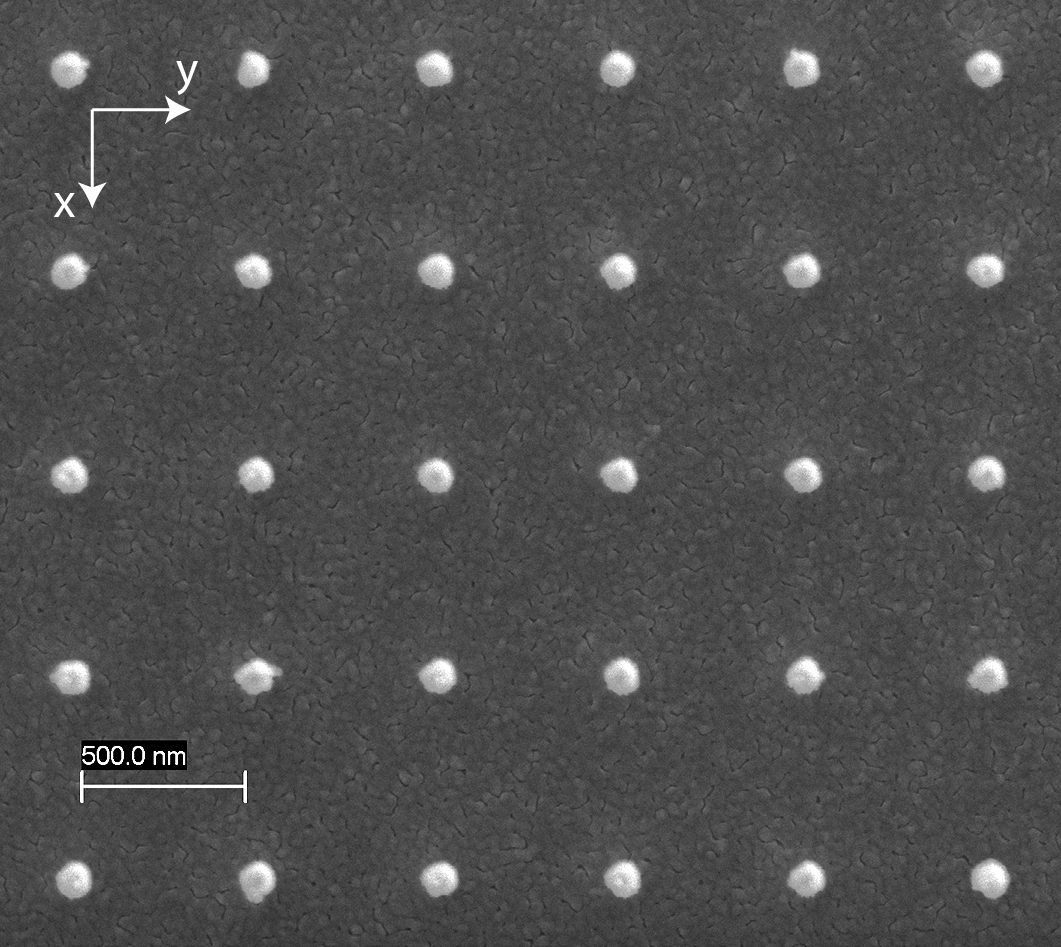}
   \caption{\textbf{Zoomed-in scanning electron microscope image of a nanoparticle array.} The same image is shown as an inset in the main text Figure~1.
  }
\end{figure}

\begin{figure}[ht!]
  \centering
    \includegraphics[width=1.0\textwidth]{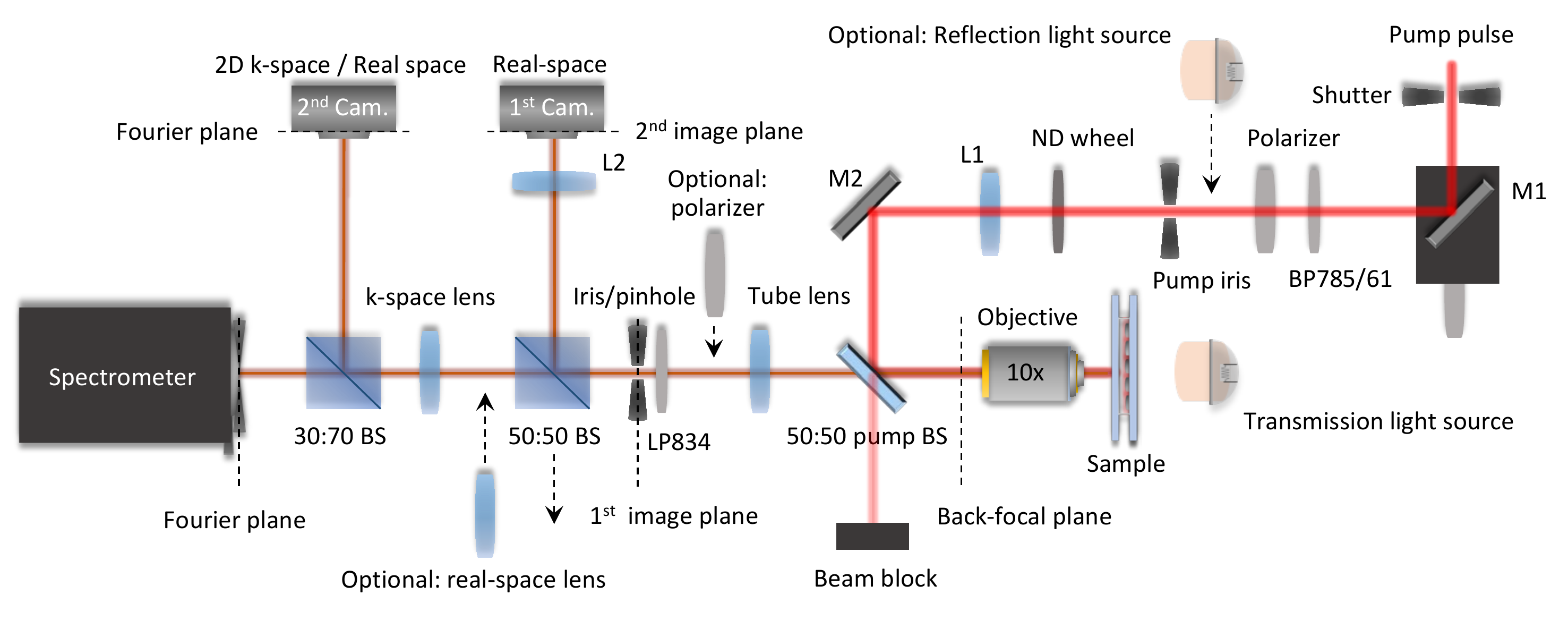}
   \caption{\textbf{Schematic of the experimental setup.} The setup allows acquiring the real space, $k$-space, and spectral information simultaneously, and it is modular to measuring transmission, reflection, and the luminescence properties of the sample. Here, BP~785/61 refers to a band-pass filter with a bandwidth of 61~nm centered at 785~nm, LP~834 refers to a long-pass filter with a cutoff at 834~nm, M stands for mirror, L for lens, ND for neutral density, and BS stands for a beam splitter with a marked fraction of (R:T).   
  }
\end{figure}

%\begin{figure}[ht!]
%  \centering
%    \includegraphics[width=1.0\textwidth]{Supp_FigXX_singleshot.pdf}
%   \caption{\textbf{Real space images for different integration times.} Real space images at %the three pump fluence regimes discussed in the manuscript: 0.96~mJcm$^{-2}$ (left column), %2.02~mJcm$^{-2}$ (center column), and 3.21~mJcm$^{-2}$ (right column). The top row (a-c) %shows real space images for a single pump pulse and the bottom row (d-f) for integrating %over 10 pump pulses.
%  }
%\end{figure}

 \begin{figure}[ht!]
  \centering
    \includegraphics[width=1.0\textwidth]{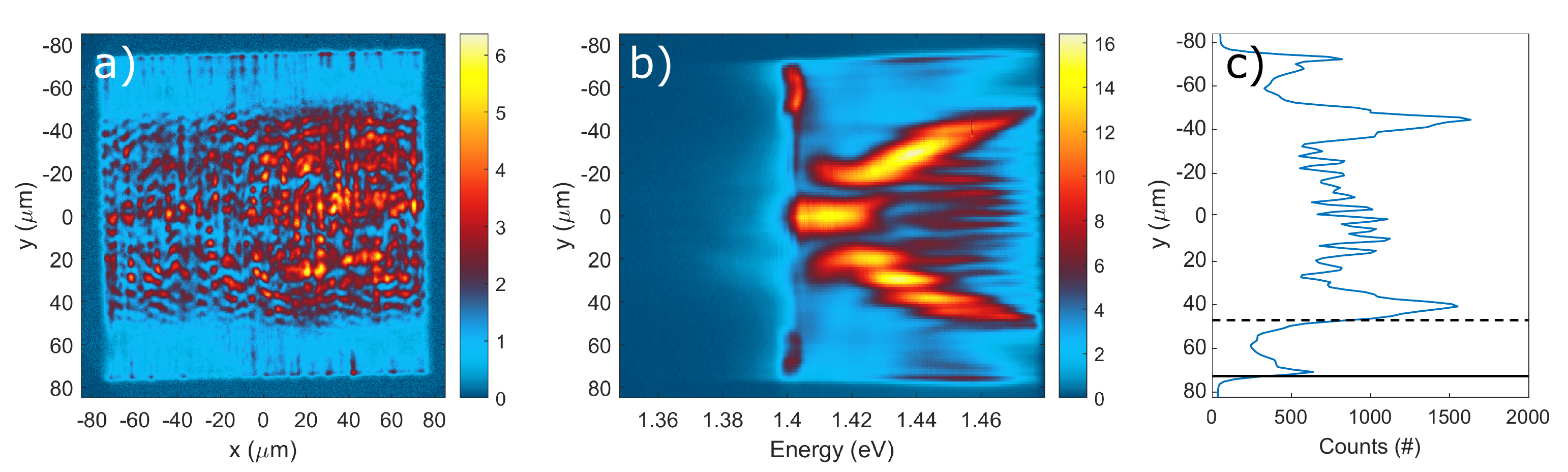}
   \caption{\textbf{Definition of the location where red shift begins.} (a) Real space image, (b) real space spectrum, and (c) integrated intensity along the x-axis of the real space spectrum at intermediate pump fluence (2.2~mJcm$^{-2}$) for 150$\times$\SI{150}{\micro m^2} lattice. Integration for the line spectrum is done over the high-energy range where the red shift begins (1.45$-$1.48~eV). We define the width of the dark zone as the distance between the array edge and the edge of the rising intensity in the real space image. This rising intensity edge corresponds to the location where the red shift begins in the real space spectrum (b). The array edge (solid line) and the edge of the rising intensity (i.e. starting point of red shift; dashed line) are defined at the half-maximum points of the rising intensity curve in (c). The distance between the solid and the dashed line is \SI{25}{\micro m}.
  }
\end{figure}

\begin{figure}[ht!]
  \centering
    \includegraphics[width=1.0\textwidth]{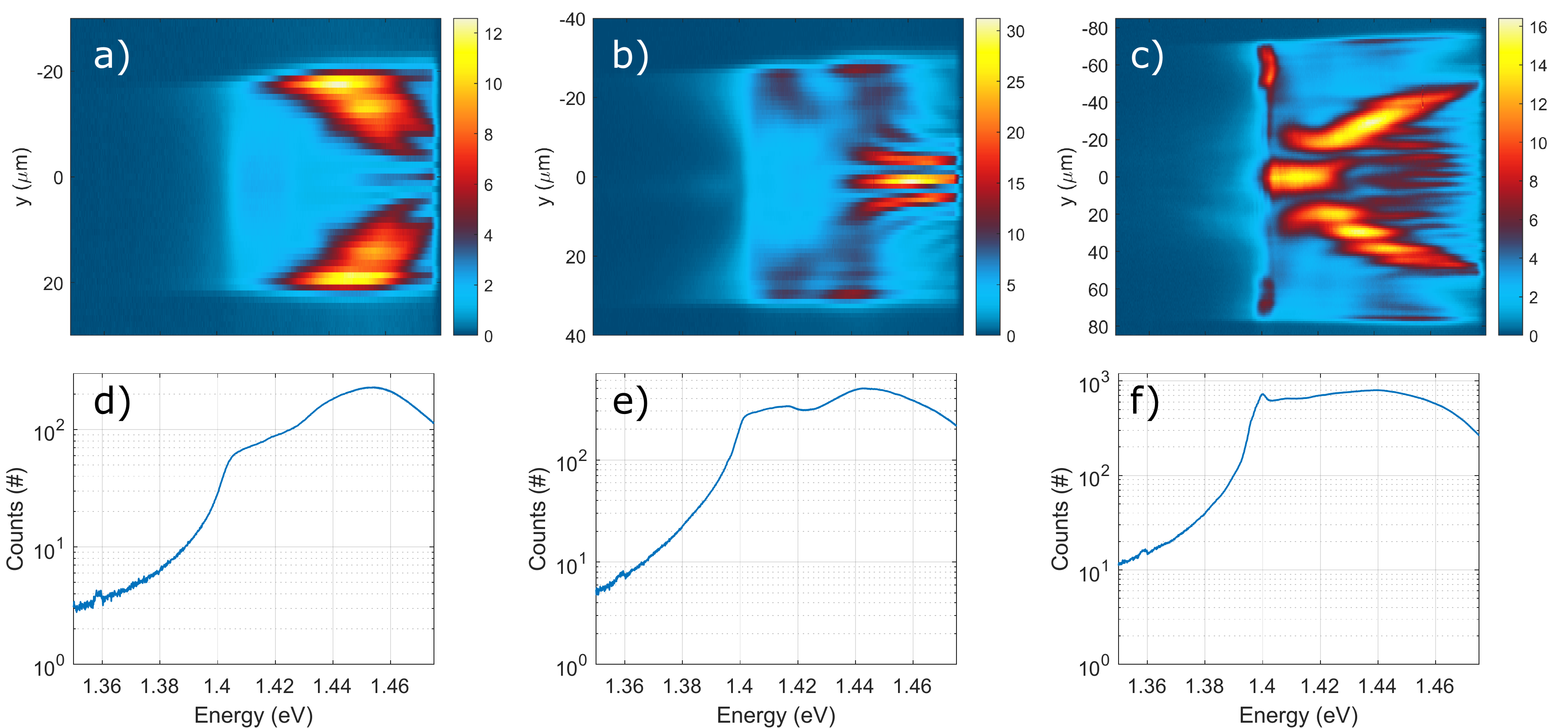}
   \caption{\textbf{Line spectra for different lattice sizes.} Real space spectra (top row) and the corresponding line spectra (bottom row) at intermediate pump fluence (2.2~mJcm$^{-2}$) for lattice sizes (a,d) 40$\times$\SI{40}{\micro m^2}, (b,e) 60$\times$\SI{60}{\micro m^2}, and (c,f) 150$\times$\SI{150}{\micro m^2}.
  }
\end{figure}

\clearpage
\begin{table}
% \small
% \footnotesize
 \scriptsize
 \centering
 \caption{\textbf{ Absorption and emission peak locations, maximum dissolving concentration, and notes on different dye molecules tested in lasing and condensation experiments.} If the absorption and emission wavelengths are measured, the used concentration (in the 1:2 DMSO:BA mixture) is given in parenthesis, otherwise a reference to the literature value is given. The column 'Max.~$c_{dye}$' indicates the maximum concentration that has been successfully dissolved in the 1:2 DMSO:BA mixture. It may not be the highest possible value, but a note is given in the last column, if higher concentrations have been unsuccessfully tried. In addition, we have written down a mark how well the dye molecule has worked in lasing experiments (similar to the ones, e.g., in~\cite{zhou_lasing_2013,Yang2015,hakala_lasing_2017,daskalakis_ultrafast_2018}).  We have compared the quantum yield (QY) of the molecules at different concentrations, and found that for IR-140, the QY at 25~mM drops to 34\% of the QY at 1~mM (highest efficiency). 
 For IR-792 the QY at 25~mM drops to 6\%, for IR-783 to 4\%, and for IR-806 to 1.4\%. 
 At 100~mM, for IR-792 the QY drops to 1.4\%, for IR-783 to 0.7\%, and for IR-806 to 0.4\%, respectively. IR-783 seems like a good candidate for the lasing experiments as its emission spectrum does not shift (or broaden) much when the concentration is increased, its QY stays decent as a function of concentration ($>$25~mM), and it emits something even at 280~mM (0.16\% of the QY at 1~mM). Unfortunately, it bleaches quickly under optical excitation. All FEW dyes~\cite{few_chemicals}, particularly FEW S0260, appear as good candidates. However, IR-792 was chosen for this (and our previous~\cite{hakala_bose-einstein_2018}) studies as it was the first tested emitter that worked nicely in the lasing and condensation experiments at high concentrations.}
 
 \begin{center}
 \begin{tabular}{l | l | l | l | l}
  & Abs. peak (nm) & Em. peak (nm) & Max. $c_{dye}$ & Notes \\ \hline
 Rhodamine 6G 		& 530 \cite{omlc_rhodamine}	& 552 \cite{omlc_rhodamine}	& 100~mM 	& Good (at visible $\lambda$). \\ % Good laser dye; Might dissolve at even at higher $c_{dye}$.
 DCM  				& 468 \cite{sigma_aldrich} 	& 627 \cite{sirah_laser}	& 40~mM 	& Good (at visible $\lambda$). \\
 IR-140 perchl.		& 835 (0.1~mM) 				& 872 / 897 (1 / 25~mM) 	& 25~mM 	& Good, insoluble at higher $c_{dye}$. \\
 IR-792 perchl.		& 811 (150~mM)			 	& 845 / 858 (1 / 200~mM) 	& 200~mM 	& Good. \\ % 5.64% and 1.37% 
 IR-780 perchl.		& 780 \cite{sigma_aldrich} 	& 834 / 844 (1 / 200~mM) 	& 200~mM 	& Product discontinued. \\ % ; QY drops at 25~mM to 6\% compared to 1~mM 
 IR-780 iodide		& 780 \cite{sigma_aldrich} 	& -- 						& -- 		& Dissolves very poorly. \\
 IR-783  			& 801 (0.1~mM) 				& 829 / 840 (1 / 280~mM) 	& 280~mM 	& Bad, bleaches quickly. \\ % 6.19 2nd day, 2.53% 1st day => mean 4.36%; and 0.70%; 280mM 0.158%
 IR-806  			& 827 (0.1~mM) 				& 856 / 865 (1 / 100~mM) 	& 100~mM 	& Bad, bleaches quickly.  \\ % 1.37%
 IR-820  			& 820 \cite{sigma_aldrich} 	& 881 (100~mM) 				& 100~mM 	& OK, not thoroughly tested. \\
 FEW S0094			& 813 \cite{few_chemicals}	& 875 (100~mM) 				& 100~mM 	& OK. \\
 FEW S0260 			& 816 \cite{few_chemicals}	& 867 / 877 (15 / 100~mM)	& 330~mM 	& Good. \\
 FEW S0712 			& 819 \cite{few_chemicals} 	& 881 (100~mM) 				& 200~mM 	& OK. \\ % 877 (50~mM)
 Styryl 9M 			& 584 \cite{sigma_aldrich} 	& 815 \cite{sirah_laser}	& 40~mM 	& OK, insoluble at higher $c_{dye}$. \\
 Perylene Red		& 578 \cite{kremer_pigmente}& 613 \cite{kremer_pigmente}& -- 		& --
 \label{tab:dyes} 
 \end{tabular}
 \end{center}
\end{table}

%\clearpage

\clearpage
%\clearpage
\subsection*{Supplementary Note 1. Effect of the band-edge energy}

We demonstrate the effect of the band edge location in Supplementary Figure~1 with two examples where the band edge is tuned either below or above the optimal energy of 1.40~eV. When the periodicity is set small so that the band edge of the $x$-polarized TE mode is at high energy (Supplementary Figure~1a-c), the propagation and red shift occurs along the lower dispersion branch of the TE mode, i.e., the red shift is not halted at the band edge. In contrast, when periodicity is set large so that the band edge resides at low energy (Supplementary Figure~1d-f), we observe the red shift of polaritons propagating along the upper dispersion branch of the TE mode but no condensation. The red-shifting population simply does not reach the band edge, which is at too low energy. The stripes in the real space spectra, Supplementary Figure~1b,e (also visible in main text Fig~2d,e and Figure~4c-f), arise from standing waves caused by counter-propagating modes. We found that the wavelength of the intensity oscillations is $\lambda_{RS}(E) = \pi/k(E)$. Comparing the $k$-space and real space spectra shows in Supplementary Figure~1a-b that the oscillations are denser at lower energies because the lower energy corresponds to larger $k$. In contrast in Supplementary Figure~1d-e, the oscillations are sparser at lower energies because the lower energy corresponds to smaller $k$. 

Interestingly while in previous studies using a dye molecule bath for photons~\cite{klaers_bose-einstein_2010,schmitt_thermalization_2015} or lattice plasmons~\cite{hakala_bose-einstein_2018} condensation required matching the lowest state energy with the energy where the molecule absorption vanishes, here that condition is not needed but the condensate formation is controlled by the interplay of the lattice size, periodicity and thermalization speed.

\subsection*{Supplementary Note 2. High luminescence signal of the condensate}

As mentioned in the main text, we obtain roughly 5 orders of magnitude stronger signal compared to the first BEC in a plasmonic lattice~\cite{hakala_bose-einstein_2018}. The increase of the signal is attributed to stimulated processes involved in our present experiments, and differences in the sample as well as the pump and detection geometry. In the previous study~\cite{hakala_bose-einstein_2018}, the pump spot overlapped only partially with the nanoparticle array, and the condensate was detected over a small part of a long array. The excited molecules emitted photons to propagating SLR modes, and the propagation took place in part of the array where there were only ground state molecules. This means that most of the excitations were lost during the propagation. 
In the present work, we pump over the whole array and also collect the luminescence from the whole array. Since the pump spot covers the whole array, all the propagation of excitations takes place in an area where there are excited molecules. The increased amount of excitations leads to a stimulated thermalization and condensation process which makes the excitations couple out as light in a very short time instead of decaying through the loss channels of the system. Since stimulated processes are involved, the increased amount of excitations leads to an output emission that is enhanced in a nonlinear manner. Furthermore, the samples that we use in the present work are more persistent towards photobleaching due to a very thick layer of dye solution (see Methods for details), enabling using higher pump fluences than in the previous work. 

%\clearpage
\subsection*{Supplementary Note 3. $T$-matrix simulation of lattice modes}

To unveil the origin of the multiple modes, we have performed multiple-scattering $T$-matrix simulations of infinite and finite arrays of cylindrical nanoparticles, with the periods in $x$ and $y$ directions as well as the nanoparticle dimensions corresponding to our system. In the $T$-matrix approach, the scattering properties of a single nanoparticle are first described in terms of vector spherical wave functions (VSWFs), giving the $T$-matrix of the particle at a given frequency. For an individual particle, the nontrivial elements of the $T$-matrix are given by solving the scattering problem of the single particle.  Next, the interactions between nanoparticles at different positions are expressed in terms of translation operators between the VSWFs with different origins \cite{xu_efficient_1998}. In case of infinite arrays, applying the Bloch boundary conditions, the electromagnetic response of a periodic nanoparticle array can be described with a matrix equation of the form 
\begin{equation}
\left( I - TW \right) a = T p_\mathrm{ext}
    \label{TMproblem}
\end{equation}
where $T=T(\omega)$ is the single particle $T$-matrix, $W(\omega, \mathbf{k})$ is a lattice sum of the VSWF translation operators, $a$ is a vector containing the coefficients of VSWFs scattered from a nanoparticle, and $p_\mathrm{ext}$ is a vector of incoming VSWF coefficients, describing the external fields driving the array. Lattice modes are defined as the solutions of Eq.~\eqref{TMproblem} with the right hand side set to zero (physically this means the waves propagate without the need of external driving) and exist only for such $(\omega,\mathbf{k})$ pairs for which the matrix $\left( I - TW \right)$ is singular (giving the dispersion relation of the array), which is equivalent to some singular value of the matrix being equal to zero. For finite arrays, the procedure is similar, but with different boundary conditions, as explained in \cite{hakala_lasing_2017,necada2020}. In practice, the particles are lossy, hence the frequency $\omega$ needs to be complex if the wave vector $\mathbf k$ is real.
We also exploit the symmetries of the system and evaluate the equation \eqref{TMproblem} separately for each irreducible representation of the little group corresponding to a given $\mathbf k$ vector. This gives us additional a priori information about the multipole polarizations of the particles in different modes.
For a more detailed description of the method, see the Supporting Information of Ref. \cite{guo_lasing_2019}, and Ref.~\cite{necada2020}.

The blue dots in Supplementary Figure~2 show the singular values of $\left( I - TW \right)$ for cylindrical nanoparticles in an infinite array. Three distinct singular values appear at the $\Gamma$-point ($\mathbf{k}=0$), while a large number of possible modes remain degenerate. This splitting into three modes comes from the finite size and cylindrical shape of the nanoparticles; in the empty lattice case all the modes are degenerate and we have seen by simulations that they remain essentially degenerate for spherical nanoparticles with less than 50 nm radius. Supplementary Figure~2 shows both the real and imaginary parts of the three distinct energies.

In finite lattices, the discrete translational symmetry of the infinite lattice is broken and the degeneracies present in the infinite case are further lifted. This is evident from Supplementary Figure~2 where the real and imaginary parts of the mode energies are shown for two different finite lattice sizes. A much larger number of distinct energies appears. The number of the modes and their energy separations and loss rates depend on the size of the lattice in a complicated way. The finite lattice size does not produce any simple "particle in a box" type distribution of the modes, since the finite size interplays with the internal multipolar modes of the nanoparticles in a non-trivial manner.
Note that the number of modes found by the method is limited by the area in the real and imaginary frequency space from which the eigenvalues are searched. For the simulations in Supplementary Figure~2, the origin and the axes (real and imaginary parts of the energy) of the contour from within the eigenvalues are searched were $(1.431$eV$+0.00i)$ and $(0.050$eV$, 0.050i)$, respectively. 

In the simulations of the finite size system, the discrete translational symmetry of an infinite system cannot be utilized and the required computational time is significant and increases with the system size. Therefore we are not able to simulate arrays of the same sizes as used in the experiments. However, based on the results of Supplementary Figure~2, one can make the qualitative conclusions that an infinite lattice has three distinct energies and finite size lattices further ones, with energy splittings that are sensitive to the lattice size and properties of the nanoparticles. The energy splittings obtained by the simulations are of the same order of magnitude as the distances between the sub-peaks in the condensate in Figs.~2a and 3j of the main text.

%\clearpage
\subsection*{Supplementary Note~4. Michelson interferometer experiment and Fourier analysis}

The fringe contrast in the interfered images is extracted with a Fourier analysis. 
First, we need to find the period of the interference fringes arising due to coherence of $E(\boldsymbol{y})$ and $E(\boldsymbol{-y})$, and set a Fourier filter for spatial frequencies accordingly. The fringe period and the corresponding spatial frequency is determined by the incoming angle of the interfered images in the experimental setup. After that, the image data is gone through column by column, inside the region of interest, and a Matlab inbuilt Fast-Fourier Transform algorithm is performed to each pixel column at a time. In the spatial frequency spectrum, we find the peak value inside a predefined frequency bandwidth and compare that to the noise floor. The peak value needs to be above a chosen threshold value. If the peak is above the threshold, the rms-sum of the frequency components within the predefined bandwidth is compared to the background level (DC value of the frequency spectrum). Finally the contrast at certain pump fluence is taken as the mean value of the contrast along the columns inside the region of interest. The threshold value for a peak-acceptance level is adjusted so that the mean contrast value found in the reference cases (incoherently summed real space image) stays below 5\%. Supplementary Figure~4 shows an explanatory example where the method is not applied for each pixel column separately but the intensity along $y$-axis averaged over $x$ (from Supplementary Figure~3b), and the intensity along $x$-axis averaged over $y$ (from Supplementary Figure~3d).

The stripes and fringes visible in Supplementary Figure~3 originate from properties of the nanoparticle arrays as well as from the Michelson intereferometer experiment. Let us first discuss the stripes specific to nanoparticle arrays. In our case one-dimensional lasing originates from linearly polarized dipolar nanoparticles. The nanoparticles are polarized in x-direction and predominantly radiate in y-direction. The polarization direction in nanoparticle array lasing is typically determined by the polarization of the pump beam, likely due to its coupling to the single particle resonance of the nanoparticle \cite{hakala_lasing_2017,wang_band-edge_2017}. The feedback is provided by counter-propagating optical modes in y-direction, and long-range radiative coupling in x-direction is not efficient. This produces one-dimensional vertical stripes in the real space images (see Supplementary Figure~3a,c) (these can be understood as individual (or just a few neighbouring) nanoparticle chains lasing independently). 
Due to the one directional coupling, one dimensional lasing shows high spatial coherence only in the direction of the feedback.

Horizontal stripes in Supplementary Figure~3a are interference fringes that arise from overlapping two real space images, one of which is flipped with respect to the x-axis, at the camera sensor in the Michelson interferometer setup. There horizontal stripes (fringes) occur on top of the vertical lasing stripes. In Supplementary Figure~3c, the flipping is done with respect to the y-axis and since there is almost no spatial coherence in the x-direction, no additional fringes are obtained (confirmed by the Fourier analysis on the amplitude of spatial frequencies explained above).

In the condensation regime, there is a more uniform intensity pattern visible in the central part of the array and the Michelson interferometer produces the interference fringes in both x- and y-direction (see Supplementary Figure~3b,d). This is in contrast to the lasing regime that shows the interference fringes only in the y-direction (Supplementary Figure~3a). Note that with increasing the size of the nanoparticles, multipolar excitation of individual nanoparticles \cite{hakala_lasing_2017,de_giorgi_interaction_2018} becomes possible and can facilitate two dimensional spatial coherence since the particle with multipole excitation can efficiently radiate in the two directions of the lattice plane. However, the nanoparticles in our current experiment (d = 100~nm) are too small (with respect to the wavelength range of interest) to exhibit 2D coherence in the lasing regime.

Finally, it is important to note that the Michelson interference fringes occur with a fixed period determined by the experimental setup (incoming angle of the overlapping images), and therefore these fringes can be distinguished from any other stripes in the real space images  (with different period).

%\clearpage
%\clearpage
\subsection*{Supplementary Note~5. Rate-equation simulation of a stimulated-emission pulse}

We use a standard four-level model for the gain medium to simulate the stimulated emission pulse when the four-level system is originally in its ground state, and excited with a 50~fs pump pulse.
The levels are labeled as follows: the pump excites the system from the level 0 to 3, there is a non-radiative decay from 3 to 2, and emission to the cavity mode is from 2 to 1. The model shows the same temporal evolution after the pump pulse as a gain-switched laser, or a $Q$-switched laser after the $Q$-switch is opened~\cite{siegman_lasers_1986}. The transition lifetimes used for the four-level gain medium are: $\tau_{32}$ = $\tau_{10}$ = 50~fs and $\tau_{21}$ = $\tau_{20}$ = 500~ps, which are similar to those used in the literature for organic dye molecules~\cite{zhou_lasing_2013,hakala_lasing_2017,daskalakis_ultrafast_2018}. 
The spontaneous emission coupling factor ($\beta$--factor) is set to $\beta = 0.001$ and the cavity lifetime to $\tau_{cav} = 100$~fs (corresponding to a typical lifetime of an SLR mode). 
The model is defined with the following coupled rate-equations, as in Ref.~\cite{daskalakis_ultrafast_2018}:
\begin{align}
\frac{dn_{ph}}{dt} &= \beta n_{ph}\frac{(N_2 - N_1)}{\tau_{21}} + \beta\frac{N_2}{\tau_{21}} - \frac{n_{ph}}{\tau_{cav}} \label{rateEq_nph} \\[15pt]
\frac{dN_0}{dt} &= -rN_0 + \frac{N_2}{\tau_{20}} + \frac{N_1}{\tau_{10}} \\[15pt]
\frac{dN_3}{dt} &= rN_0 - \frac{N_3}{\tau_{32}} \\[15pt]
\frac{dN_2}{dt} &= -\beta n_{ph}\frac{(N_2 - N_1)}{\tau_{21}} - \frac{N_2}{\tau_{21}} - \frac{N_2}{\tau_{20}} + \frac{N_3}{\tau_{32}} \\[15pt]
\frac{dN_1}{dt} &= \beta n_{ph}\frac{(N_2 - N_1)}{\tau_{21}} + \frac{N_2}{\tau_{21}} - \frac{N_1}{\tau_{10}},
\end{align}
where the populations of each level are denoted with $N_i$ and the transition lifetimes with $\tau_i$. Here, $n_{ph}$ is the photon number in the mode (in our case the number of polaritons). 
Parameter $r$ is the pump rate proportional to the pump intensity that has a Gaussian temporal shape. 

In the model, the threshold value for population inversion is defined by comparing the gain and loss terms for the photon number in Eq.~\eqref{rateEq_nph}. The optical gain must overcome the loss, and at the threshold they are equal
\begin{equation}
\beta n_{ph}\frac{(N_2 - N_1)}{\tau_{21}} = \frac{n_{ph}}{\tau_{cav}}.
\end{equation}
With a definition of $N^*=N_2 - N_1$, the threshold value becomes
\begin{equation}
N^*_{th} = \frac{\tau_{21}}{\beta \tau_{cav}}. \label{eq:threshold_value}
\end{equation}

%\clearpage
\subsection*{Supplementary Note~6. Description of the dissipative quantum model}

We have studied the thermalization mechanism qualitatively with a microscopic quantum model including multiple cavity modes coupled to a single two-level system that is coupled to a shifted harmonic oscillator that describes the rotational-vibrational degrees of freedom within a molecule. The results of the model are presented in Supplementary Figure~5.

The system is described by the Holstein-Tavis-Cummings model \cite{tavis_exact_1968,cwik_polariton_2014,herrera_cavity-controlled_2016,strashko_organic_2018,wu_when_2016} with the Hamiltonian ($\hbar=1$)
\begin{align}
\label{totalhamiltonian}
H =  \sum_i \omega_i a_i^\dag a_i + \frac{\omega_m}{2}\sigma^z + \sum_i \left( g a_i \sigma^+  + g^* a^\dag_i \sigma^-  \right) + \omega_v b^\dag b + \omega_v \sqrt{S}(b^\dag + b)\sigma^z.
\end{align}
Here $a_i^\dag$ is the bosonic creation operator of the cavity mode of index $i$, $\sigma^z$ and $ \sigma^\pm$ are the Pauli operators describing the two-level structure of the molecule and $b^\dag$ is the bosonic creation operator corresponding to the vibrational mode of the molecule. Furthermore, $\omega_k$ is the energy of the cavity mode of index $i$, $\omega_v$ is the energy of the vibrational mode, $\omega_m$ is the energy of the two-level system, $g$ is the coupling between cavity modes and the molecule, and $S$ is the Huang-Rhys parameter. Rotating wave approximation has been used in the simulation, assuming that the coupling $|g|$ is significantly smaller than the molecule and cavity mode frequencies. This is true with the parameters used in the simulation: $\omega_i = {1.40, 1.41, 1.42, 1.43, 1.44, 1.45}$, $\omega_m=1.45$, and $g=0.0075$ (in eV). Other parameters in the Hamiltonian used in the simulation are: $\omega_v=0.03$~eV and $S=0.1$.

Dissipations of an open quantum system are taken into account in the Lindblad formalism, which yields the master equation~\cite{breuer_theory_2002}:
\begin{align}
\label{mastereqf}
\frac{\delta \rho(t)}{\delta t} = & \frac{i}{\hbar}[\rho(t),H] + \sum_i \kappa_i\mathcal{L}[a_i] + \gamma_m\mathcal{L}[\sigma^-] + \gamma_z\mathcal{L}[\sigma^z] \nonumber \\ 
&+ \gamma_{v,+}\mathcal{L}[b^\dag-\sqrt{S}\sigma^z] + \gamma_{v,-}\mathcal{L}[b-\sqrt{S}\sigma^z],
\end{align}
where $\mathcal{L}[O] = O \rho O^\dag - \frac{1}{2}O^\dag O \rho - \frac{1}{2}\rho O^\dag O $ is the Lindblad superoperator, $\kappa$ is the cavity dissipation rate, $\gamma_m$ and $\gamma_z$ are the radiative dissipation and dephasing rates of the molecule, and $ \gamma_{v,\pm}$ are the rates that describe thermal excitation ($+$) and dissipation ($-$) of the vibrational mode. These rates are $\gamma_{v,+}=\gamma_v n_B$ and $\gamma_{v,-}=\gamma_v (n_B + 1)$, where $n_B=1/[\mathrm{exp}(\omega_v/k_\mathrm{B} T_\mathrm{mol}-1)]$ is the occupation probability of the vibrational mode of energy $\omega_v$ at thermal equilibrium. Solving Eq.~\eqref{mastereqf} for $<a_i^\dag a_i>$ , $<\sigma^z>$, and $<b^\dag b>$ gives the time evolution of occupation of the cavity modes and the molecule as well as excitation of the vibrational mode. Parameters used in the simulation for Supplementary Figure~5 are $\gamma_v=4\times10^{-3}$, $\gamma_m=1\times10^{-6}$, $\gamma_z=1\times10^{-3}$, $\kappa=6\times10^{-3}$, and $T_\mathrm{mol} = 25\times10^{-3}$ (in eV). The simulation was performed using Python 3 with QuTiP toolbox~\cite{johansson_qutip_2012}. We use this model to illustrate a possible mechanism for the observed thermalization process, however, quantitative comparison to our experiments is not meaningful due to the simplicity of the model.

%\clearpage
\subsection*{Supplementary Note~7. Different pump pulse durations}

We studied the condensation phenomenon as a function of pump pulse duration and found that thermalization and condensation happens only for sub-250~fs pump pulses. Comparison of 50~fs and 500~fs pump pulses is presented in Supplementary Figure~6. It is evident that the longer excitation pulse results in only one (lasing) threshold. The distributions at around the threshold (blue and red curves in Supplementary Figure~6a,e are similar with both pulse durations, but at the higher fluences (yellow, purple, and green curves) the distributions are very different. With a 500~fs pulse, the population does not reach a thermal Maxwell--Boltzmann distribution, and no narrow peaks appear at the band edge. Besides the luminescence intensity, the different threshold behaviour is clearly visible in the FWHM curves (Supplementary Figure~6c,g). The FWHM is significantly decreased with both pulse durations at the first (lasing) threshold but only in the 50-fs case, the FWHM is decreased even further at the second (condensation) threshold. Note that at intermediate pump fluences, the 50~fs pulse shows a sharp increase of the FWHM because the maximum of the line spectra is found at higher energies (see Supplementary Figure~6d) as the thermalizing population dominates the signal. Examples of real space and $k$-space images and spectra for the 500~fs are shown in Supplementary Figure~7, for pump fluences corresponding to the (a,c) lasing threshold and (b,d) beyond the condensation threshold of the 50~fs pump pulse. Importantly the real space and $k$-space images and spectra look nearly identical for both pump fluences for the 500~fs pulse duration, confirming that there indeed is no second threshold where the condensation would take place.

\subsection*{Supplementary Note~8. Estimation of polariton-polariton interaction strength}

As mentioned in the main text, lasing and condensation take place at higher energy than the band edge of the lower polariton branch. Since the whole dispersion blue shifts as a function of pump fluence, the blue shift may be associated to degradation of strong coupling rather than e.g. Coulombic interactions. However, such saturation-caused non-linearity can also be considered as effective polariton-polariton interaction~\cite{carusotto_quantum_2013}. 

To make a rough estimation of the strength of such interaction based on the observed blue shift, we use a mean-field approximation estimate for low polariton density (linear regime)~\cite{Deng2010}. %(cite Deng et.al., Reviews of Modern Physics 82, 2010).
In this case the blue shift $\Delta E$ is linearly dependent on the polariton density $n$ and the interaction constant $g$, $\Delta E = g n$. Putting in our experimentally observed values of $\Delta$E $\sim$ 20 meV and $10^{17}$ polaritons/m$^2$ we obtain a value of \textsl{g} $\sim$ 0.2$\mu$eV$\mu$m$^2$. This value can be converted into a dimensionless interaction strength $\tilde{g}=g m / \hbar^2$ which can be defined for a two-dimensional system\cite{bloch_many-body_2008, radonjic_interplay_2018}. The polariton mass is estimated as in our previous work~\cite{hakala_bose-einstein_2018} by fitting a parabola to the band edge of the lower polariton dispersion branch. Fitting to both TM and TE modes of the coupled system gives an estimate of the effective mass in range $10^{-37}...10^{-35}$~kg. Using these values and the $g$ calculated above, the dimensionless interaction strength $\tilde{g}=g m / \hbar^2$ is of the order of $10^{-7}...10^{-5}$.

\clearpage

\bibliography{bec_paper2019}
\bibliographystyle{ieeetr}

\end{document}